\documentclass[12pt]{article}
\pdfoutput=1
\usepackage{putex}
\usepackage{graphicx}
\usepackage{caption}
\usepackage{subcaption}
\usepackage{epstopdf}
\usepackage{enumerate}
\usepackage{cite}
\usepackage{tensor}
\usepackage{slashed}
\usepackage[aligntableaux=center]{ytableau}
\usepackage[utf8]{inputenc}
\usepackage{multirow}

\newcommand{\abs}[1]{\left\lvert #1 \right\rvert}

\newcommand {\be} {\begin {equation}}
\newcommand {\ee} {\end {equation}}

\newcommand {\bes} {\begin {equation*}}
\newcommand {\ees} {\end {equation*}}

\newcommand{\es}[2] {\begin{equation} \label{#1} \begin{split} #2 \end{split} \end{equation}}

\ytableausetup
{mathmode, boxsize=1em}

\newcommand{\cO}{{\cal O}}

\newcommand{\beq}{\begin{equation}}
\newcommand{\eeq}{\end{equation}}

\def\<{\langle}
\def\>{\rangle}

\begin{document}

\preprint{PUPT-2501}

\institution{PU}{Joseph Henry Laboratories, Princeton University, Princeton, NJ 08544, USA}

\title{Anomalous dimensions of scalar operators in QED$_3$}

\authors{Shai M.~Chester and Silviu S.~Pufu}

\abstract {
The infrared dynamics of $2+1$ dimensional quantum electrodynamics (QED$_3$) with a large number $N$ of fermion flavors is governed by an interacting CFT that can be studied in the $1/N$ expansion. We use the $1/N$ expansion to calculate the scaling dimensions of all the lowest three scalar operators that transform under the $SU(N)$ flavor symmetry as a Young diagram with two columns of not necessarily equal heights and that have vanishing topological charge.  In the case of $SU(N)$ singlets, we study the mixing of $(\bar \psi_i \psi^i)(\bar \psi_j \psi^j)$ and $F_{\mu\nu} F^{\mu\nu}$, which are the lowest dimension parity-even singlets.  Our results suggest that these operators are irrelevant for all $N>1$.
}

\date{}

\maketitle

\tableofcontents
\setlength{\unitlength}{1mm}

\newpage
\section{Introduction and summary}
\label{INTRO}	 
Quantum electrodynamics in 2+1 dimensions (QED$_3$) with $N$ (two-component complex) charged fermions can be shown to flow to an interacting CFT perturbatively in the $1/N$ expansion \cite{Appelquist:1988sr,Nash:1989xx}.   In this CFT, there are many quantities that have been computed to various orders in $1/N$.  Examples include: the scaling dimensions of the lowest $SU(N)$ singlet scalars \cite{Gracey:1993iu,Gracey:1993sn,Rantner:2002zz,2008PhRvB..78e4432X}, adjoints \cite{Hermele05, Kaul:2008xw}, and a couple of other scalar operators \cite{2008PhRvB..78e4432X};  the scaling dimensions of monopole operators \cite{Borokhov:2002ib,Pufu:2013vpa,Dyer:2013fja}; the two-point functions of the canonically-normalized stress-energy tensor and of the conserved currents \cite{Huh:2013vga,Huh:2014eea,Giombi:2016fct};  the $S^3$ free energy \cite{Klebanov:2011td}; as well as various finite temperature quantities \cite{Kaul:2008xw}.\footnote{See also \cite{DiPietro:2015taa,Giombi:2015haa,Chester:2015wao} for estimates of some of the same quantities coming from the $4-\epsilon$ expansion.}  Our goal here is to add to this list the scaling dimensions of many more operators:  For each $SU(N)$ irreducible representation with two columns of fixed lengths, we identify the $3$ lowest-lying scalar operators with zero monopole charge, and we compute their scaling dimensions to order $1/N$.  

Our interest in the scaling dimensions of scalar operators transforming in non-trivial representations of $SU(N)$ comes in part from the recent conformal bootstrap study \cite{Chester:2016wrc} of QED$_3$.  This study focused on unitarity and crossing symmetry constraints on the four-point function of monopole operators carrying a single unit of topological charge.  The OPE of such a monopole operator and its conjugate contains operators transforming under the flavor $SU(N)$ as precisely the irreps considered in this paper, namely two-column irreps of the form
\es{OPEreps}{
SU(N) \text{   irreps:   } \qquad   N-n \left\{ \rule{0pt}{1.65cm} \right. \begin{ytableau}
 {}& {}  \\
{} & {}  \\
\none[\vdots] & \none[\vdots] \\
{} & {}  \\
{}  \\
\none[\vdots]\\
{}
\end{ytableau}\raisebox{.62cm}{$\left. \rule{0pt}{.95cm} \right\}n$} = \left(1^{N-2n},2^n\right) \,,\qquad n=0\,,1\,,\dots\,,N/2\,,
}
where $\left(\lambda_1^{\nu_1},\lambda_2^{\nu_2},\dots\right)$ denotes a Young tableau with $\nu_i$ rows of length $\lambda_i$, and the $n=1$ case is the adjoint.  In tensor notation, the irrep \eqref{OPEreps} can be represented as a traceless tensor with $n$ antisymmetric fundamental indices and $n$ antisymmetric anti-fundamental indices.  While  the bootstrap study \cite{Chester:2016wrc} only examined relatively small values of $N$ (namely $N=2, 4$, and $6$), future studies may be able to access larger values of $N$, and in order to assess their accuracy, one would benefit from more large $N$ analytical approximations than those currently available in the literature.   We thus develop the large $N$ expansion for the scaling dimensions of scalar operators transforming as \eqref{OPEreps} under $SU(N)$.  Of course, such large $N$ expansions could also be useful independently of the conformal bootstrap program, for instance if one engineers a new material that exhibits critical behavior described by QED$_3$ with a sufficiently large number of flavors.  

An additional motivation exists for computing the scaling dimensions of lowest-lying parity even $SU(N)$ singlet operators.  As we explain below, the lowest such operator has scaling dimension approximately equal to 4 at $N = \infty$, with negative $1/N$ corrections.   If this operator becomes relevant at some finite value of $N$, it may completely change the IR physics if no tuning is performed.  It is conceivable that for $N \leq N_\text{crit}$, the deep IR corresponds to a chiral symmetry breaking phase and that $N_\text{crit}$ can be estimated from when the scaling dimension of the lowest lying $SU(N)$ singlet approaches~$3$~\cite{DiPietro:2015taa, 2008PhRvB..78e4432X,Giombi:2015haa}.  Computing this scaling dimensions as a function of $N$ would allow us to estimate~$N_\text{crit}$.

Let us present a summary of our results. For any $n\geq 0$, for which the $SU(N)$ irrep is given by \eqref{OPEreps}, we denote the lowest dimension operator by ${\cal O}_{n}$.  As we show in Section~\ref{numOps}, $SU(N)$ group theory requires that for $n>0$, ${\cal O}_{n}$ must be constructed from a product of precisely $n$ distinct fermions anti-symmetrized in their $SU(N)$ indices and symmetrized in their spinor indices and a product of $n$ distinct anti-fermions with the same property.  Furthermore, only a single operator can be built in this way, namely\footnote{The construction of this operator requires $n \leq N/2$. The regime where $n$ is comparable to $N$ is outside of the validity range of our approximation---we first fix $n$ and then take $N$ to be large.}
\es{kOpMain}{
  (\cO_n)^{i_1\ldots i_n}{}_{i_{n+1} \ldots i_{2n}}=&\psi^{[i_1}_{(\alpha_{1}}\dots\psi^{i_n]}_{\alpha_{n)}}\bar\psi_{[i_{n+1}}^{(\alpha_{1}}\dots\bar\psi_{i_{2n}]}^{\alpha_{n})} - \text{($SU(N)$ traces)}
}
where $\alpha_m = 1, 2$ are Lorentz spinor indices and $i_m = 1, \ldots, N$ are flavor indices---see Section~\ref{review} for our conventions. This operator is parity even (odd) depending on whether $n$ is even (odd). We provide a formula for the scaling dimension $\Delta_n$ of this operator to order $1/N$ in Eq.~\eqref{NOpFinal} for all $n>0$.  This formula is rather complicated, so we record the scaling dimensions here only for the first several cases:\footnote{The scaling dimension $\Delta_1$ was already computed in \cite{Hermele05, Kaul:2008xw}.  $\Delta_2$ agrees with the result of Sections~II.B and II.C of \cite{2008PhRvB..78e4432X}.  The other operator in Section~II.C of \cite{2008PhRvB..78e4432X}, with scaling dimension $4 + \frac{64}{3 \pi^2 N} + O(N^{-2})$, is a four-fermion operator transforming under $SU(N)$ as the irrep $(2^{N-1}, 4^1)$.}
\es{246}{
&\Delta_1=2-\frac{64}{3\pi^2N}+O(1/N^2)\,, \qquad
\Delta_2=4-\frac{64}{\pi^2N}+O(1/N^2) \,, \\
&\Delta_3=6-\frac{128}{\pi^2N}+O(1/N^2) \,, 
 \qquad \Delta_4=8-\frac{640}{3\pi^2N}+O(1/N^2) \,, \\
 & \text{etc.}
}

Next, we consider the lowest dimension operator in the same $SU(N)$ irrep as $\cO_n$ but with opposite parity. For that purpose, we must consider an operator constructed with one more $\psi$ and $\bar\psi$ each than $\cO_n$. As we will show, for $0<2n<N$ there are two linearly independent such operators, which can be taken to be\footnote{The construction of ${\cal O}_n'$ requires $n \leq N/2 - 1$, and the construction of ${\cal O}_n''$ requires $n \leq N/2$.  The regime where $n$ is comparable to $N$ is outside of the validity range of our approximation---we first fix $n$ and then take $N$ to be large.}
\es{kOp2Main}{
 ( \cO_n')^{i_1\ldots i_n}{}_{i_{n+1} \ldots i_{2n}} &=\frac{1}{\sqrt{N}}\sum_{k=2n+1}^N\psi^{[i_1}_{(\alpha_{1}}\dots\psi^{i_n}_{\alpha_{n}}\psi^{k]}_{\alpha_{n+1)}}\bar\psi_{[i_{n+1}}^{(\alpha_{1}}\dots\bar\psi_{i_{2n}}^{\alpha_{n}}\bar\psi_{k]}^{\alpha_{n+1})}-\text{($SU(N)$ traces)}\,,\\
  ( \cO_n'')^{i_1\ldots i_n}{}_{i_{n+1} \ldots i_{2n}} &=(\cO_n)^{i_1\ldots i_n}{}_{i_{n+1} \ldots i_{2n}} \frac{\bar\psi_i^\alpha \psi^i_\alpha }{\sqrt{N}}\,.
}
 By considering the mixing of these two operators \eqref{kOp2Main}, we calculate the scaling dimensions $\Delta_{n,\pm}'$ to order $1/N$ for all $n>0$. Since the final expression (Eq.~\eqref{knOPs}) is rather complicated, we will only record here the scaling dimensions for the first several cases: 
\es{2462}{
&\Delta_{1,\pm}'=4+\frac{8\left(25\pm\sqrt{2317}\right)}{3\pi^2N}+O(1/N^2)\,,
  \qquad\Delta_{2,\pm}'=6+\frac{32\pm160}{\pi^2N}+O(1/N^2)\,,\\
&\Delta_{3,\pm}'=8+\frac{4\left(-21\pm\sqrt{19189}\right)}{3\pi^2N}+O(1/N^2)\,,\qquad\Delta_{4,\pm}'=10+\frac{64\left(-26\pm\sqrt{2362}\right)}{15\pi^2N}+O(1/N^2)\,,\\
&\text{etc.}
}

Lastly, the case $n=0$ ($SU(N)$ singlet) requires special treatment.  The lowest dimension parity odd $SU(N)$ singlet is $\cO_0=\frac{1}{\sqrt{N}}\bar\psi_i^\alpha \psi^i_\alpha$. Its scaling dimension is \cite{Rantner:2002zz}
 \es{DeltaSinglet}{
   \Delta_{0} = 2 + \frac{128}{3 \pi^2 N} + O(1/N^2)  \,.
 }
The two lowest dimension parity even operators are mixtures of the operators $\left(\bar\psi_i\psi^i\right)\left(\bar\psi_j\psi^j\right)$ and $F_{\mu\nu} F^{\mu\nu}$.  We find that the scaling dimensions are
 \es{ScalingSinglet}{
 \Delta'_{0,\pm} =4 + \frac{64 (2 \pm \sqrt{7}  ) }{3 \pi^2} \frac 1N + O(1/N^2) \,.
 }
This result agrees with that of Ref.~\cite{2008PhRvB..78e4432X} that was obtained through a different method.\footnote{Ref.~\cite{2008PhRvB..78e4432X} studied the mixing of the operators $\left(\bar\psi_i\psi^i\right)\left(\bar\psi_j\psi^j\right)$ and $\left(\bar\psi_i\gamma_\mu\psi^i\right)\left(\bar\psi_j\gamma^\mu\psi^j\right)$ by adding these operators to the action and studying the renormalization of their couplings as one integrates out momentum shells.  In our approach, we extract the scaling dimensions from the matrix of two-point functions, and in doing so we can make use of the equations of motion.  The gauge field equation of motion, $\bar\psi_i\gamma_\mu\psi^i= 0$, implies that the two-point function of $\left(\bar\psi_i\gamma_\mu\psi^i\right)\left(\bar\psi_j\gamma^\mu\psi^j\right)$ vanishes at separated points. Instead of considering the mixing of $\left(\bar\psi_i\psi^i\right)\left(\bar\psi_j\psi^j\right)$ and $\left(\bar\psi_i\gamma_\mu\psi^i\right)\left(\bar\psi_j\gamma^\mu\psi^j\right)$, we consider the mixing of $\left(\bar\psi_i\psi^i\right)\left(\bar\psi_j\psi^j\right)$ and $F_{\mu\nu} F^{\mu\nu}$, as we do in Section~\ref{4singlet}.  Despite the different methods, we obtain the same result as Ref.~\cite{2008PhRvB..78e4432X}.  It would be interesting to perform a similar computation to the one in this paper in the case of an $SU(2)$ gauge theory and compare with the results of \cite{2008PhRvB..78e4432X}.} Extrapolating \eqref{ScalingSinglet} to finite $N$, one finds that all parity-even $SU(N)$ singlets are  irrelevant for all values of $N>1$.  This result might suggest that the interacting CFT obtained in the $1/N$ expansion extends to all values of $N>1$, in agreement with the recent lattice simulations of \cite{Karthik:2015sgq}.\footnote{See, however, \cite{Strouthos:2008kc} where it was observed that for $N=2$ there is spontaneous chiral symmetry breaking.  Also, the $F$-theorem \cite{Giombi:2015haa} implies that chiral symmetry breaking is ruled out for $N\geq 10$.}  It is worth mentioning that the scaling dimension of the four-fermion parity-even singlet was also estimated from the $4-\epsilon$ expansion in \cite{DiPietro:2015taa}, where it was found that this operator is irrelevant only for $N>2$ \cite{DiPietro:2015taa}.  It would be interesting to understand how the mixing between the four-fermion operator and $F_{\mu\nu}^2$ studied here affects the $4-\epsilon$ expansion estimates.

The rest of this paper is organized as follows.  In Section~\ref{review}, we set up our conventions and Feynman rules for QED$_3$. Sections~\ref{kferm} and \ref{4singlet} contain the bulk of our computations.

\section{Setup and conventions}
\label{review}

Before we delve in the computations of the various scaling dimensions mentioned above, let us describe our conventions and the setup of our computation.  The Euclidean signature Lagrangian of QED$_3$ with $N$ fermion flavors is
 \es{LEuc}{
  {\cal L} = \frac{1}{4e^2} F_{\mu\nu} F^{\mu\nu} - \bar \psi_i \gamma^\mu (\partial_\mu + i A_\mu) \psi^i  \,,
 }
where $e$ is the gauge coupling.  The gamma matrices obey the Clifford algebra $\{\gamma^\mu, \gamma^\nu\} = 2 \delta^{\mu\nu} I$ and can be taken to be equal to the Pauli matrices $\gamma^\mu = \sigma_\mu$, for $\mu = 1, 2, 3$.  We choose to write fundamental spinor indices as lower and fundamental $SU(N)$ indices as upper, as in $\psi^i_\alpha$, with $i=1, \ldots, N$ and $\alpha = 1, 2$.  Anti-fundamental indices have the opposite index placement, as in $\bar \psi_i^\alpha$.  In the following we try to avoid as much as possible writing down explicit spinor indices, but we do write down the $SU(N)$ flavor indices explicitly.   Repeated indices are always summed over.

As will become clear shortly, the gauge coupling $e$ drops out of all computations in the IR CFT\@.  Therefore, one can think of the fermions in \eqref{LEuc} as carrying any gauge charge, and not necessarily the smallest unit of charge allowed by the $U(1)$ gauge symmetry.  The results of this paper are thus independent of the gauge charge of the fermions.

\subsection{Derivation of Feynman rules}

\subsubsection{Feynman rules with standard gauge fixing}

A slightly cumbersome but natural option is to work with the Feynman rules derived directly from the Lagrangian \eqref{LEuc} supplemented by the standard gauge fixing term
 \es{GaugeFixing}{
  {\cal L}_\text{gauge fixing} = -\frac {1}{2e^2} \frac 1{\xi} (\partial_\mu A^\mu)^2 \,.
 }
In momentum space, the fermion propagator $G(p)$ and the gauge field propagator $D_{\mu\nu} (p)$ are 
 \es{GFerm}{
  G(p) = \frac{i \gamma_\mu p^\mu}{p^2} \,, \qquad
  D_{\mu\nu}^\text{Max} (p) = \frac{e^2}{p^2} \left(\delta_{\mu\nu} - (\xi + 1) \frac{p_\mu p_\nu}{p^2} \right) \,.
 }
 \begin{figure}[t]
\begin{center}
\includegraphics[width = 1\textwidth]{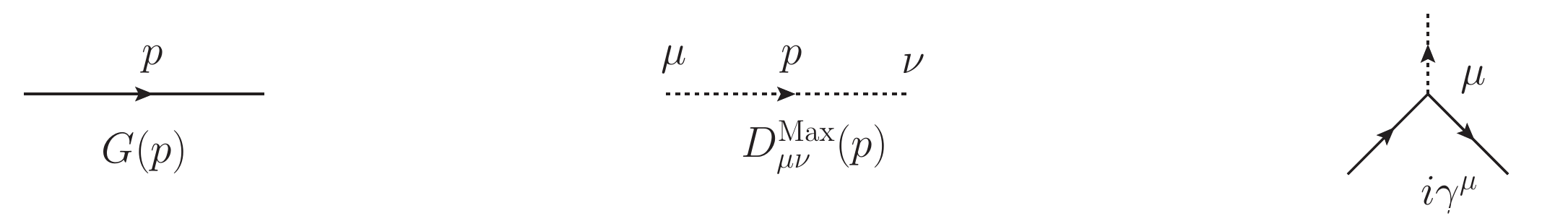}
 \caption{Feynman rules with standard gauge fixing. \label{FeynmanMaxwell}}
 \end{center}
 \end{figure}
The gauge-fermion vertex factor is simply $i \gamma^\mu$.  See Figure~\ref{FeynmanMaxwell}.   Computing diagrams using these rules is then straightforward.  The IR CFT behavior can be extracted by taking the limit of small external momenta.  This limit is equivalent to taking $e^2 \to \infty$ in all correlation functions because by dimensional analysis $e^2$ always appears as $e^2/\abs{p}$, where $p$ is one of the external momenta.  

Using the Feynman rules in Figure~\ref{FeynmanMaxwell} is cumbersome for two reasons.  The first reason is that at the CFT fixed point the Maxwell term is irrelevant, so there should be a way of performing the computation such that $e^2$ never appears and no limit needs to be taken at the end.  In other words, there should be a way of performing the computation where the $e^2 \to \infty$ limit is taken from the very beginning.  The second reason is that at each order in $1/N$ there is an infinite number of fermion bubble diagrams that always get resummed in the same way, so one should resum them once and for all.

\begin{figure}[t]
\begin{center}
\includegraphics[width = 1\textwidth]{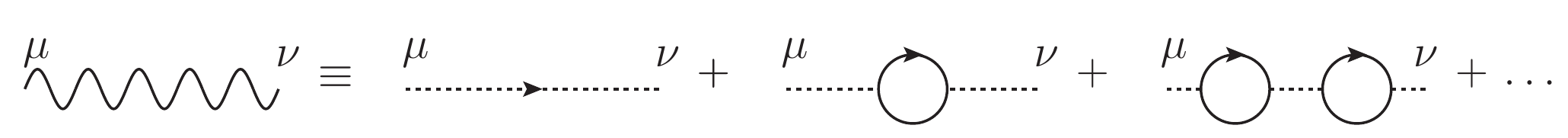}
 \caption{Effective photon propagator defined as sum of fermion bubble diagrams.\label{EffectivePropag}}
 \end{center}
 \end{figure}

Let us address the second concern first.  In order to avoid resumming the same bubble diagrams every time, one can define an effective gauge field propagator obtained after the resummation.  See Figure~\ref{EffectivePropag}.  In order to obtain an explicit expression for the effective gauge propagator, it is convenient to work with the position-space fermion Green's function obtained by Fourier transforming~\eqref{GFerm}:
 \es{psiTwoPoint}{
   \langle \psi^{i}(x) \bar \psi_{j}(y) \rangle_\infty
     = \delta^i_j\, G(x, y) \,, \qquad
     G(x, y) = \int \frac{d^3 p}{(2 \pi)^3} G(p) e^{-i p \cdot (x - y)} = \frac{ \gamma_\mu (x^\mu - y^\mu)}{4 \pi \abs{x-y}^3} \,.
 }
Each fermion bubble is nothing but the two-point function of the gauge current $j^\mu = \bar \psi_i \gamma^\mu \psi^i$;  in position space, it is
 \es{GotPiPhi}{
  \Pi^{\mu\nu}(x, y) &=   \langle j^\mu(x) j^\nu(y) \rangle_\infty =  - \frac{N}{8 \pi^2 \abs{x}^6} \left( \delta^{\mu\nu} x^2 - 2 x^\mu x^\nu \right) \,,
 }
as follows from performing the required Wick contraction and using \eqref{psiTwoPoint}.   Passing to momentum space, one has
 \es{PiMom}{
   \Pi^{\mu\nu}(x, y) = \int \frac{d^3p}{(2 \pi)^3} \Pi^{\mu\nu} (p) e^{-i p \cdot (x - y)} \,, \qquad \Pi^{\mu\nu}(p)=  \frac{N \abs{p}}{16} \left(\delta^{\mu\nu} - \frac{p^\mu p^\nu}{p^2} \right)  \,,
 }
as follows from the formulas given in \eqref{FT4}.  As defined above, the effective gauge field propagator is just the sum of the fermion bubbles and takes the form of a geometric series:
 \es{DEff}{
  D^\text{eff}_{\mu\nu}(p) &= D_{\mu\nu}^\text{Max}(p) - D_{\mu\rho}^\text{Max}(p) \Pi^{\rho\sigma}(p) D_{\sigma\nu}^\text{Max}(p) + \ldots  \\
   &= - \frac{\xi e^2 p_\mu p_\nu}{p^4}  + \frac{16}{N \abs{p}}  \left( \delta_{\mu\nu} - \frac{p_\mu p_\nu}{p^2} \right) + O(p^2/e^2)  \,.
 }
One can thus replace the gauge propagator \eqref{GFerm} with \eqref{DEff} in order to not have to resum the bubble diagrams every time, and otherwise compute Feynman diagrams as usual. 
 
\subsubsection{Feynman rules with non-standard gauge fixing}

As already mentioned, it would be nice to have a way of performing computations at the CFT fixed point without having to carry around $e^2$ and to take the limit $e^2 \to \infty$ at the end of the computation.  Unfortunately, the Maxwell propagator \eqref{GFerm} and the effective propagator \eqref{DEff} do not generally have finite limits as $e^2 \to \infty$, so this limit cannot in general be taken at the beginning of the computation.

An exception occurs in the gauge $\xi = 0$, where the effective gauge propagator \eqref{DEff} does have a finite limit as $e^2 \to \infty$ and one can indeed take $e^2 \to \infty$ from the beginning.  As we now show, it is also possible to modify the gauge fixing term \eqref{GaugeFixing} so as to have a one-parameter family of gauge-fixing terms, not just that for $\xi = 0$, for which one can take $e^2 \to \infty$ from the beginning.

Instead of \eqref{GaugeFixing}, one can consider the non-local gauge-fixing term 
 \es{GaugeFixingAgain}{
  \tilde {\cal S}_\text{gauge fixing} = \frac{N}{32(\zeta-1)} \int d^3x \, \int d^3 y\,  \frac{\partial_\mu A^\mu(x) \partial_\nu A^\nu(y)}{2 \pi^2 \abs{x - y}^2}
   = \frac{N}{32 (\zeta-1)} \int \frac{d^3 p}{(2 \pi)^3}  \frac{p_\mu p_\nu A^\mu(p)  A^\nu(-p)}{\abs{p}}
 }
where $\zeta$ is a gauge-fixing parameter.  Using \eqref{GaugeFixingAgain} instead of \eqref{GaugeFixing}, the Maxwell gauge field propagator in \eqref{GFerm} gets replaced by
 \es{GaugePropTilde}{
  \tilde D_{\mu\nu}^\text{Max}(p) = \frac{e^2}{p^2} \left(\delta_{\mu\nu} - \frac{p_\mu p_\nu}{p^2} \right) 
   + \frac{16(\zeta-1) }{N} \frac{p_\mu p_\nu}{\abs{p}} + O(p^2/e^2)\,,
 }
and the effective gauge propagator in \eqref{DEff} gets replaced by
  \es{GaugePropEffTilde}{
  \tilde D_{\mu\nu}^\text{eff} (p) = \frac{16}{N \abs{p}}  \left( \delta^{\mu\nu} - \zeta \frac{p^\mu p^\nu}{p^2} \right) + O(p^2/e^2) \,.
 }
As advertised, this expression has a finite limit as $e^2 \to \infty$ for any $\zeta$.  Gauge invariant observables should of course be independent of $\zeta$.

\subsection{Summary of Feynman rules}

\begin{figure}[t]
\begin{center}
\includegraphics[width = 1\textwidth]{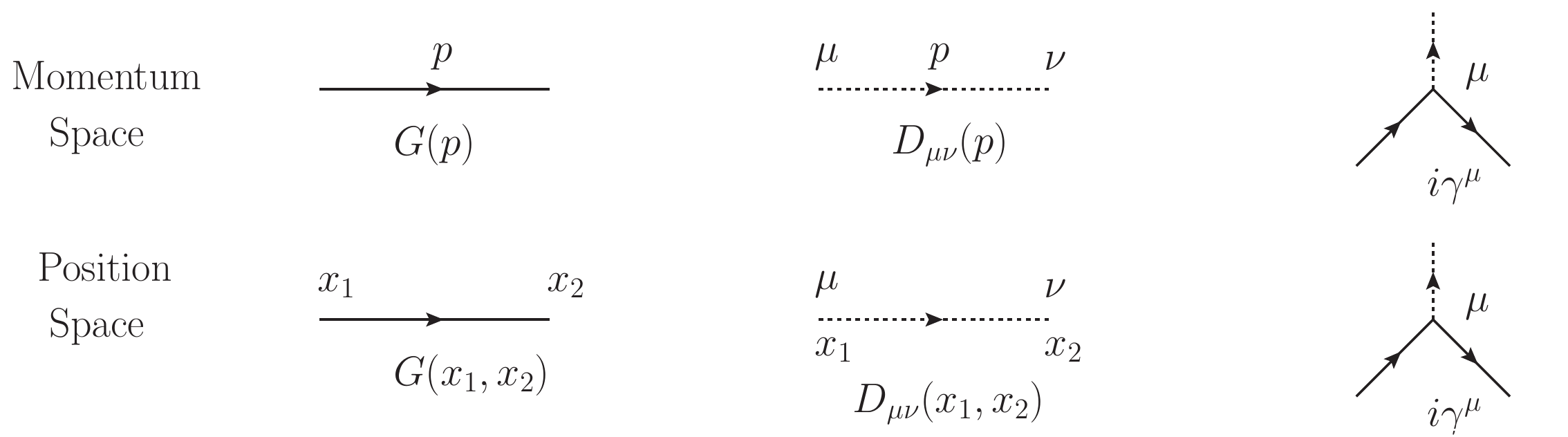}
 \caption{Feynman rules used in this paper.\label{fig:feynman}}
 \end{center}
 \end{figure}

To summarize, the momentum and position space Feynman rules we will work with are:  
 \es{Summary}{
  G(p) &= \frac{i \gamma_\mu p^\mu}{p^2}\,, \qquad D_{\mu\nu} (p) = \frac{16}{N \abs{p}}  \left( \delta_{\mu\nu} - \zeta \frac{p_\mu p_\nu}{p^2} \right)  \,, \\
  G(x_1, x_2) &=  \frac{ \gamma_\mu x_{12}^\mu }{4 \pi \abs{x_{12}}^3} \,, \qquad 
   D^{\mu\nu}(x_1, x_2) = \frac{8}{\pi^2 N \abs{x_{12}}^2} \left[ (1 - \zeta) \delta^{\mu\nu} +  2 \zeta \frac{x_{12}^\mu x_{12}^\nu }{\abs{x_{12}}^2} \right] \,,
 }
where $x_{12} \equiv x_1 - x_2$ and  the position space expression for $D_{\mu\nu}$ is derived in \eqref{ABGauge}.  The vertex factor is $i \gamma^\mu$.   See Figure~\ref{fig:feynman}.

In working with the effective gauge field propagator $D^{\mu\nu}$ one should keep in mind that this propagator stands for the sum of the bubble diagrams in Figure~\ref{EffectivePropag}, so one should not count the same Feynman diagram multiple times.  In particular, one should not consider any effective gauge propagators renormalized by fermion bubbles, for instance as on the RHS of Figure~\ref{EffectivePropag} if the dotted lines were replaced by wavy lines.

\subsection{General strategy for anomalous dimension computation}
\label{GENERALANOMALOUS}

In this paper, we compute anomalous dimensions from the matrix of two point functions in $1/N$ perturbation theory.  Suppose that there are $r$ operators ${\cal O}_a$, $a = 1, \ldots, r$ that have the same quantum numbers and scaling dimension $\Delta^{(0)}$ at leading order in $N$.  The matrix of two-point functions has a large $N$ expansion of the form
 \es{TwoPointGeneral}{
  {\cal M}_{ab}(x) \equiv \langle {\cal O}_a(x) \bar {\cal O}_b(0) \rangle 
   = {\cal M}_{ab}^{(0)}(x) + {\cal M}_{ab}^{(1)}(x) \frac{1}{N} + \ldots \,.
 }
We expect the following $x$ dependence of the first two coefficients:
 \es{DefineM}{
  {\cal M}_{ab}^{(0)}(x) = \frac{{\bf N}_{ab}}{\abs{x}^{2 \Delta^{(0)}}} \,, \qquad
   {\cal M}_{ab}^{(1)}(x) = \frac{1}{ \abs{x}^{2 \Delta^{(0)}}} \left[- {\bf M}_{ab}\log (\abs{x}^2 \Lambda^2) + O(\abs{x}^0) \right] \,.
 }
This expression serves as a definition of the $r\times r$ matrices ${\bf N}$ and ${\bf M}$.  Here, $\Lambda$ is the UV cutoff, which is required in order to make the argument of the logarithm dimenisonless.  At order $1/N$ the anomalous dimensions are the eigenvalues $\Delta^{(1)}_a$ of the matrix 
  \es{gammaDef}{
   \Delta^{(1)} = {\bf N}^{-1} {\bf M} 
  }
(see for instance \cite{Constable:2002vq}).  The total scaling dimensions are thus 
 \es{TotalScaling}{
  \Delta_a = \Delta^{(0)} + \Delta^{(1)}_a \frac 1N + O(1/N^2) \,.
 }
In the examples below, we compute the matrices ${\bf N}$ and ${\bf M}$ and use this procedure to extract~$\Delta_a$.

\subsection{Previous results}

In the following we will use the previously computed results for the leading $1/N$ corrections to the scaling dimensions of the fermion field $\psi$ and that of the 2-fermion singlet $\cO_0=\frac{1}{\sqrt{N}}\bar\psi_i\psi^i$ \cite{Hermele05, Rantner:2002zz}. 

\subsubsection{Correction to fermion propagator}

Because the gauge fixing term \eqref{GaugeFixingAgain} is conformally invariant, the two point function of $\psi$ has powerlaw decay for any $\zeta$.  However, the corresponding scaling dimension $\Delta_\psi$ will depend on $\zeta$ and does not have to obey the unitarity bound for a spin-$1/2$ operator.  We have
 \es{FermCorrel}{
  \langle \psi(x_1) \bar \psi(x_2) \rangle \propto \frac{ \gamma_\mu x_{12}^\mu}{4 \pi \abs{x_{12}}^{1 + 2 \Delta_\psi}}  \,.
 }
Expanding in $1/N$, we have $\Delta_\psi = 1 + \Delta_\psi^{(1)} \frac 1N + O(N^{-2}) $ and 
 \es{FermCorrelExpansion}{
  \langle \psi(x_1) \bar \psi(x_2) \rangle = G(x_1, x_2) \left[ 1 + \left(- \Delta_\psi^{(1)} \log (x^2 \Lambda^2) + O(\abs{x}^0) \right) \frac 1N  + O(N^{-2}) \right]  \,.
 }
 \begin{figure}[t]
\begin{center}
\includegraphics[width = 1\textwidth]{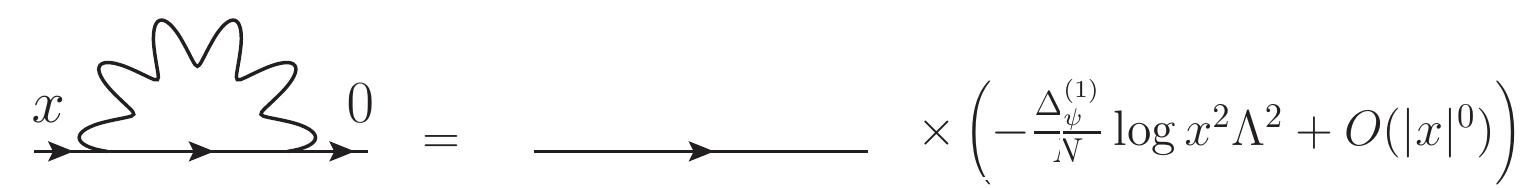}
 \caption{Diagram for $\Delta_\psi^{(1)}$.\label{FermProp}}
 \end{center}
 \end{figure}
The correction coefficient $\Delta_\psi^{(1)}$ can be found from the diagram in Figure~\ref{FermProp}.  It is found to be \cite{Rantner:2002zz}
 \es{DeltaPsi}{
  \Delta_\psi^{(1)}  = \frac{4}{\pi^2} \left( \frac 13 - \zeta \right) \,.
 }

\subsubsection{$2$-fermion singlet}
\label{SINGLET}

The dimension of $\cO_0=\frac{1}{\sqrt{N}}\bar\psi_i\psi^i$ is \cite{ Rantner:2002zz}
 \es{Delta0Large}{
  \Delta_0 = 2 + \Delta_0^{(1)} \frac 1N + O(N^{-2}) \,, \qquad \Delta_0^{(1)} = \frac{128}{3\pi^2} \,.
 } 
We exhibit the diagrams that were used in evaluating $\Delta_0^{(1)}$ in Figure~\ref{Delta0Diagrams}.

\subsubsection{$2$-fermion adjoint}
\label{TWOADJOINT}

\begin{figure}[t]
\begin{center}
\includegraphics[width = .85\textwidth]{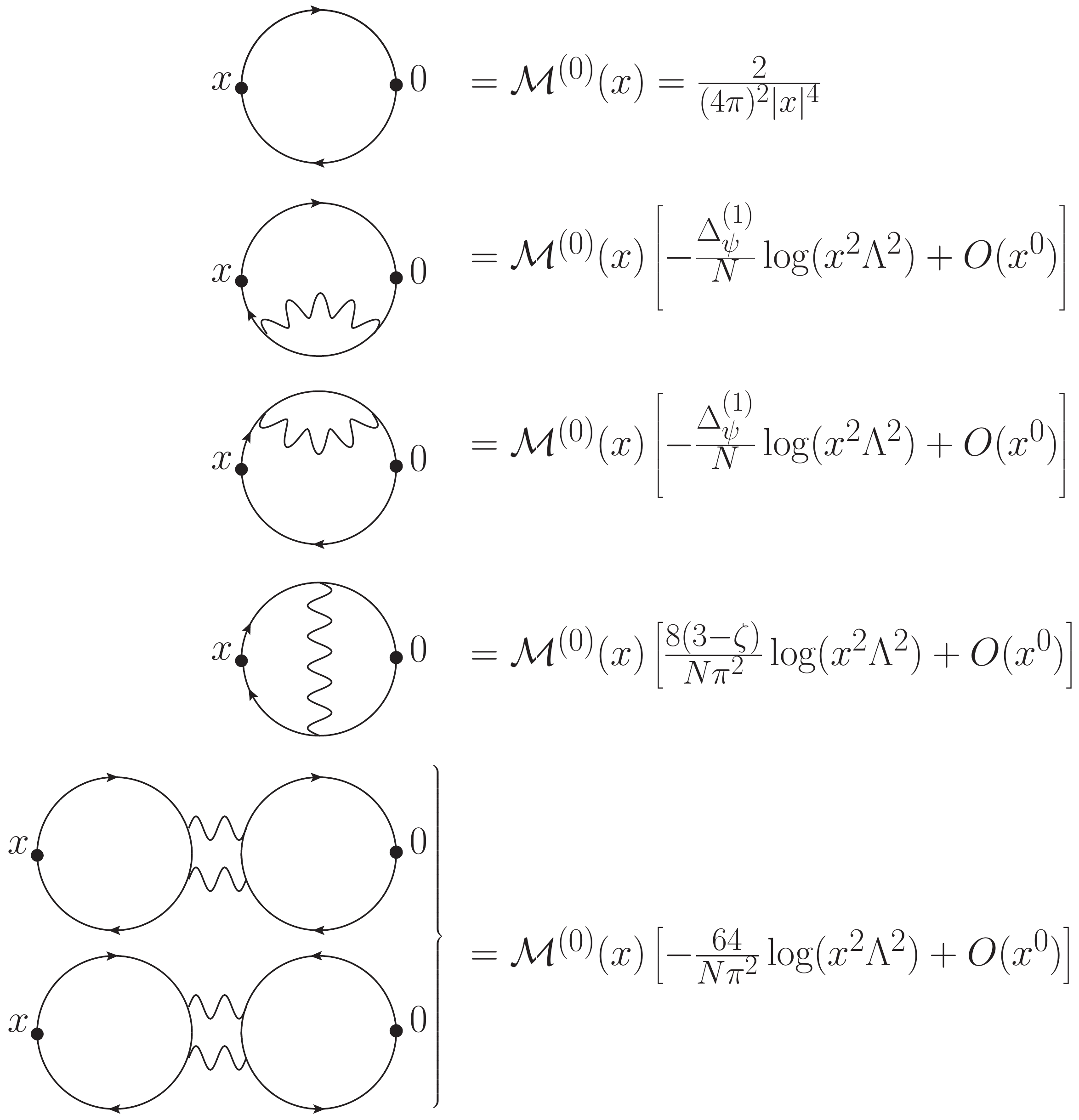}
 \caption{Diagrams for $\Delta_0^{(1)}$.\label{Delta0Diagrams}}
 \end{center}
 \end{figure}

Similarly, the dimension of the 2-fermion adjoint $\cO_1=\bar\psi_i\psi^j - \frac 1N \delta^j_i \bar\psi_k\psi^k$ is  \cite{Hermele05}
 \es{Delta1Large}{
  \Delta_1 = 2 + \Delta_1^{(1)} \frac 1N + O(N^{-2}) \,, \qquad \Delta_1^{(1)} = -\frac{64}{3\pi^2} \,.
 } 
The diagrams that contribute to $\Delta_0^{(1)}$ are the same as those in Figure~\ref{Delta0Diagrams} except for the last two.
\section{Operators in representation $\left(1^{N-2n},2^{n}\right)$}
\label{kferm}

In this section we consider the three lowest-lying scalar operators transforming in the irrep $(1^{N-2n}, 2^n)$ of $SU(N)$, with $n>0$.  For a Young diagram representation, see Eq.~\eqref{OPEreps}.  In the case $n=1$, the lowest-lying operator is the 2-fermion adjoint discussed in Section~\ref{TWOADJOINT}.

\subsection{Number of operators}
\label{numOps}

The scalar operators in $\left(1^{N-2n},2^{n}\right)$, being gauge invariant, must be constructed from an equal number of $\psi$'s and $\bar \psi$'s.  Let us count how many linearly independent operators we can construct out of $m$ $\psi$'s and $m$ $\bar \psi$'s and determine the smallest value of $m$ that is necessary in order to be able to construct at least one such operator.

\ytableausetup
{mathmode, boxsize=1.5em}
\begin{figure}[t!]
\begin{center}
  $$ \mbox{\Large $m$ } \left\{ \rule{0pt}{2.4cm} \right. \begin{ytableau}
 {}  \\
{}  \\
{}\\
\none[\vdots] \\
{}  \\
{}\\
{}\\
\end{ytableau}=\mbox{\Huge$\sum\limits_{\mbox{\large$j=\left\{ \frac m2\right\}$}}^{ \mbox{\large$\frac m2$}}$}\left(
\overbrace{\begin{ytableau}
 {} &{} &{}&\none[\hdots] & {}\\
\end{ytableau}}^{\mbox{\Large$2j$}} \,\,\mbox{\Large$,$}\, \, \mbox{\Large$\frac m2+j$} \left\{ \rule{0pt}{2.4cm} \right. \begin{ytableau}
 {}& {}  \\
{} & {}  \\
\none[\vdots] & \none[\vdots] \\
{} & {}  \\
{}  \\
\none[\vdots]\\
{}
\end{ytableau}\raisebox{1cm}{$\left. \rule{0pt}{1.45cm} \right\} \mbox{\Large$\frac m2-j$}$} \right)
$$\\
$$\mbox{\large$SU(2N)\qquad\qquad\qquad  SU(2)\qquad\qquad \qquad \qquad SU(N) \quad$}$$
\caption{Decomposition of $[\psi]^m$ under $SU(2)\times SU(N)$}\label{DECOMP}
\label{tab}
\end{center}
\end{figure}

Let us consider the $m$ $\psi$'s and the $m$ $\bar\psi$'s separately at first. The $\psi$'s transform as fundamentals both under the flavor group $SU(N)$ and the space-time group $SU(2)$.  Since there are $2N$ such $\psi$'s, we can formally combine them in a fundamental vector of a larger group $SU(2N)$, which is not a symmetry group of the theory but nevertheless a convenient bookkeeping device.  In terms of the product group $SU(2N)$, the product of $m$ $\psi$'s, denoted $[\psi]^m$, transforms as $\left(1^{m}\right)$ because the $\psi$'s are all anti-commuting. The representation $\left(1^{m}\right)$ of $SU(2N)$ decomposes under $SU(2)\times SU(N)$ as 
\es{decomp}{
[\psi]^m:\quad\left(1^{m}\right)\to\bigoplus_{j=\{m/2\}}^{m/2}\left((2j),  \left(1^{2j},2^{m/2-j}\right) \right) \,,
}
where $\{x\}$ denotes the fractional part of $x$ and $(2j)$ denotes the spin-$j$ irrep of $SU(2)$.  See Figure~\ref{DECOMP}.   The product $[\bar \psi]^m$ transforms in the representation conjugate to \eqref{decomp}:
\es{decompConj}{
[\bar \psi]^m:\quad\left(1^{m}\right)\to\bigoplus_{j=\{m/2\}}^{m/2}\left((2j),  \overline{\left(1^{2j},2^{m/2-j}\right)} \right) \,.
}

The product $[\psi]^m \times [\bar\psi]^m$ transforms in a reducible representation of $SU(2) \times SU(N)$ that can be obtained by simply multiplying \eqref{decomp} and \eqref{decompConj}.  It contains operators with spin ranging from $0$ to $m$.  The spin-0 operators appear only when multiplying a spin $j$ irrep in \eqref{decomp} with a spin $j$ irrep in \eqref{decompConj}, and so they transform under $SU(N)$ as 
\es{singlets}{
\left[[\psi]^m\times[\bar\psi]^m\right]_{\text{$SU(2)$ singlets}}=\bigoplus_{j=\{m/2\}}^{m/2}\left[\left(1^{2j},2^{m/2-j}\right)  \otimes \overline{\left(1^{2j},2^{m/2-j}\right)}\right]\,.
}

Each term in the sum \eqref{singlets} can be further decomposed as a sum of irreducible representations of $SU(N)$.  Performing this decomposition is a straightforward group theory exercise, and one can then count how many times the irrep $\left(1^{N-2n},2^{n}\right)$ we are interested in appears in this decomposition.  The result is that if $m<n$, the irrep $\left(1^{N-2n},2^{n}\right)$ does not appear at all:  we need at least $n$ $\psi$'s and $n$ $\bar \psi$'s in order to construct an operator transforming in $\left(1^{N-2n},2^{n}\right)$.  If $m=n$, the irrep $\left(1^{N-2n},2^{n}\right)$ appears in the decomposition of \eqref{singlets} only once, and it comes from the term $j=n/2$; the corresponding operator can be written explicitly as
\es{kOp}{
 (\cO_n)^{i_1\ldots i_n}{}_{i_{n+1} \ldots i_{2n}} = \psi^{[i_1}_{(\alpha_{1}}\dots\psi^{i_n]}_{\alpha_{n)}}\bar\psi_{[i_{n+1}}^{(\alpha_{1}}\dots\bar\psi_{i_{2n}]}^{\alpha_{n})} - \text{($SU(N)$ traces)} \,,
}
where we symmetrize and anti-symmetrized with unit weight, and the traces are over $SU(N)$ indices.  This operator is non-zero only for $2n \leq N$.  When $m=n+1$, the irrep $\left(1^{N-2n},2^{n}\right)$ appears in \eqref{singlets} twice, once coming from $j=m/2$ and once form $j=m/2 - 1$.  The corresponding linearly independent operators can be taken to be
\es{kOp2}{
 (\cO_n')^{i_1\ldots i_n}{}_{i_{n+1} \ldots i_{2n}} &= \frac{1}{\sqrt{N}}\sum_{k=1}^N\psi^{[i_1}_{(\alpha_{1}}\dots\psi^{i_n}_{\alpha_{n}}\psi^{k]}_{\alpha_{n+1)}}\bar\psi_{[i_{n+1}}^{(\alpha_{1}}\dots\bar\psi_{i_{2n}}^{\alpha_{n}}\bar\psi_{k]}^{\alpha_{n+1})}- \text{($SU(N)$ traces)}\,,\\
 (\cO_n'')^{i_1\ldots i_n}{}_{i_{n+1} \ldots i_{2n}} &= \frac{\left(\bar\psi_i\psi^i\right)}{\sqrt{N}} (\cO_n)^{i_1\ldots i_n}{}_{i_{n+1} \ldots i_{2n}} \,,
}
where ${\cal O}_n'$ corresponds to $j=m/2$ and ${\cal O}_n''$ is a linear combination of an operator from $j=m/2-1$ and $j=m/2$ that is easy to write down.  Note that ${\cal O}_n'$ is non-zero only if $2n<N$ and ${\cal O}_n''$ is non-zero only for $2n \leq N$.  It is straightforward to use the same method to also count the multiplicity of the irrep $\left(1^{N-2n},2^{n}\right)$ when $m \geq n+2$, but we will not be concerned with those cases here.

\subsection{Scaling dimension of ${\cal O}_n$}
\label{scalDim}
We consider a particular operator representing \eqref{kOp} by taking $i_k = k$:
\es{NOpDef}{
\cO^{}_n = \psi^{[1}_{(\alpha_1}\dots\psi^{n]}_{\alpha_n)}\bar\psi_{[n+1}^{(\alpha_1}\dots\bar\psi_{2n]}^{\alpha_n)} \,,
} 
where the trace term in \eqref{kOp} does not contribute because all the $i_k$ are distinct.  This operator can be rewritten as
\es{NOp}{
  \cO^{}_n =
    (-1)^{\frac{n(n+1)}{2}}\frac{1}{n!}\sum_{\sigma\in S_n}\text{sig}(\sigma) \cO_n^{(\sigma)}(x)\,, 
      \qquad 
   \cO_n^{(\sigma)}\equiv  \bar\psi_{n+1}\psi^{\sigma(1)}\dots\bar\psi_{2n}\psi^{\sigma(n)} \,,
} 
where the spinor indices are contracted between adjacent fermions, and $\text{sig}(\sigma)$ is the signature of the permutation $\sigma \in S_n$.  The conjugate of $\cO_n^{(\sigma)}$ is
\es{NOpconj}{
\bar\cO_n^{(\sigma)}\equiv\bar\psi_{\sigma(1)}\psi^{n+1}\dots\bar\psi_{\sigma(n)}\psi^{2n} \,.
}

We would like to express the two-point function of ${\cal O}_n$ as in Section~\ref{GENERALANOMALOUS}. Directly from the definition \eqref{NOp}, we can write 
 \es{TwoPointPermTmp}{
  \langle\cO_n(x) \bar\cO_n(0)\rangle
    =  \frac{1}{(n!)^2} \sum_{\sigma', \sigma'' \in S_n} 
   \text{sig}(\sigma') \text{sig}(\sigma'') \langle \cO_n^{(\sigma')}(x) \bar\cO_n^{(\sigma'')}(0) \rangle \,.
 }
For any permutation $\tau \in S_n$, we can perform the relabeling $\psi_i \to \psi_{\tau(i)}$, which shows that $\langle \cO_n^{(\sigma')}(x) \bar\cO_n^{(\sigma'')}(0) \rangle = \langle \cO_n^{(\sigma' \tau)}(x) \bar\cO_n^{(\sigma'' \tau)}(0) \rangle$.  We can freely apply such a transformation to each term in \eqref{TwoPointPermTmp} separately.  Taking $\tau = (\sigma')^{-1}$ and denoting $\sigma'' \tau = \sigma$, we see that \eqref{TwoPointPermTmp} reduces to
  \es{TwoPointPerm}{
  \langle\cO_n(x) \bar\cO_n(0)\rangle
    =  \frac{1}{n!} \sum_{\sigma \in S_n} 
   \text{sig}(\sigma) \langle \cO_n^{(I)}(x) \bar\cO_n^{(\sigma)}(0) \rangle \,,
 }
where $I$ is the identity permutation.  Noticing that  each permutation in a given conjugacy class gives an equal contribution to the two-point function, we can express \eqref{TwoPointPerm} as a sum over conjugacy classes $C_{n,i}$ of the symmetric group $S_n$:
\es{N2p}{
  \langle\cO_n(x) \bar\cO_n(0)\rangle = \frac{1}{n!}\sum_{C_{n,i}\in\text{Cl}(S_n)} \text{sig}(C_{n,i})|C_{n,i}|\langle\cO_n^{(I)}(x) \bar\cO_n^{(C_{n,i})}(0)\rangle \,.
}

Since conjugacy classes of the symmetric group $S_n$ will appear several times in this section, let us briefly review their properties.  Conjugacy classes of $S_n$ are in one-to-one correspondence with integer partitions of $n$.  Suppose we write such an integer partition corresponding to a conjugacy class $C_{n,i}$ as 
 \es{Partition}{
  n = \sum_{j=1}^n a_{ij} j \,,
 }
for some positive integers $a_{ij}$.  All permutations in $C_{n,i}$ have $a_{ij}$ cycles of length $j$.  In terms of this data, the size and signature of $C_{n,i}$ can be expressed as
\es{conjSize}{
|C_{n,i}| = \frac{n!}{\prod_{j=1}^n(j)^{a_{ij}}(a_{ij}!)} \,, \qquad 
  \text{sig}(C_{n,i})  =  (-1)^{\sum_{j=1}^n a_{ij} (j-1) } \,.
}
See Table~\ref{vals}.  
\begin{table}[htp]

\begin{center}
   \begin{tabular}{c||c|c|c|c|c}
    $n$ & $i$ & partition for $C_{n, i}$ & $a_{i}$ & $\abs{C_{n, i}}$ & $\text{sig}(C_{n, i})$   \\
    \hline
    1 & 1 & 1  & $\begin{pmatrix} 1 \end{pmatrix}$ & 1 & 1   \\
    \hline
     \multirow{2}{*}{$2$} & 1 & $2$ & $\begin{pmatrix} 0 & 1 \end{pmatrix}$ & $1$ & $-1$   \\
     & 2 & $1+1$ & $\begin{pmatrix} 2 & 0 \end{pmatrix}$ & $1$ & $1$   \\
    \hline 
    \multirow{3}{*}{$3$} & 1 & $3$ & $\begin{pmatrix} 0 & 0 & 1 \end{pmatrix}$ & $2$ & $1$   \\
     & 2 & $2+1$ & $\begin{pmatrix} 0 & 1 & 1 \end{pmatrix}$ & $3$ & $-1$   \\
     & 3 & $1+1+1$ & $\begin{pmatrix} 3 & 0 & 0 \end{pmatrix}$ & $1$ & $1$   \\
    \hline 
    \multirow{5}{*}{$4$} & 1 & $4$ & $\begin{pmatrix} 0 & 0 & 0 & 1 \end{pmatrix}$ & $6$ & $-1$   \\
     & 2 & $3+1$ & $\begin{pmatrix} 1 & 0 & 1 & 0 \end{pmatrix}$ & $8$ & $1$   \\
     & 3 & $2+2$ & $\begin{pmatrix} 0 & 2 & 0 & 0 \end{pmatrix}$ & $3$ & $1$   \\
     & 4 & $2+1+1$ & $\begin{pmatrix} 2 & 2 & 0 & 0 \end{pmatrix}$ & $6$ & $-1$   \\
     & 5 & $1+1+1+1$ & $\begin{pmatrix} 4 & 0 & 0 & 0 \end{pmatrix}$ & $1$ & $1$   \\                
   \end{tabular}
\end{center}

\caption{Conjugacy class data for $n=1\,,2\,,3\,,4$.}
\label{vals}
\end{table}%

\subsubsection{Leading order}

At leading order at large $N$, we can evaluate $\langle\cO_n^{(I)}(x) \bar\cO_n^{(C_{n,i})}(0)\rangle$ using Wick contractions with the propagator in \eqref{Summary}.   Since each permutation cycle of length $j$ contributes  $- \tr \left[G(x, 0) G(0, x) \right]^j$, we have 
 \es{OsOsleading}{
  \langle\cO_n^{(I)}(x) \bar\cO_n^{(C_{n,i})}(0)\rangle_{(0)} 
    &=   \prod_{j=1}^n \left( - \tr \left[G(x, 0) G(0,x) \right]^j \right)^{a_{ij}}
     = \text{sig}(C_{n,i}) \frac{2^{\sum_{j=1}^n a_{ij}}}{(4\pi)^{2n}|x|^{4n}} \,,
 }
where we used  \eqref{conjSize} and the fermion propagator in \eqref{Summary}.  Then, using \eqref{N2p}, we find
\es{NOpleading}{
\langle\cO_n(x) \bar\cO_n(0)\rangle_{(0)}
=&\frac{1}{n!}\sum_{C_{n,i}\in\text{Cl}(S_n)}|C_{n,i}|\left(\frac{2^{\sum_{j=1}^n a_{ij}}}{(4\pi)^{2n}|x|^{4n}}\right) \,.
}
 The sum $\sum_{j=1}^n a_{ij}$ gives the number of cycles in conjugacy class $C_{n,i}$. Explicitly,
\es{NOpleading246}{
\langle\cO^{}_1(x) \bar\cO^{}_1(0)\rangle_{(0)}=&\frac{1}{8\pi^2|x|^2} \,, \\
\langle\cO^{}_2(x) \bar\cO^{}_2(0)\rangle_{(0)}=&\frac{1}{2(4\pi)^4|x|^8}\left(4+2\right)=\frac{3}{256\pi^4|x|^8} \,, \\
\langle\cO^{}_3(x) \bar\cO^{}_3(0)\rangle_{(0)}=&\frac{1}{6(4\pi)^6|x|^{12}}\left(8+12+4\right)=\frac{1}{1024\pi^6|x|^{12}} \,,
}
and so on.

 \begin{figure}[t]
\begin{center}
\includegraphics[width = .5\textwidth]{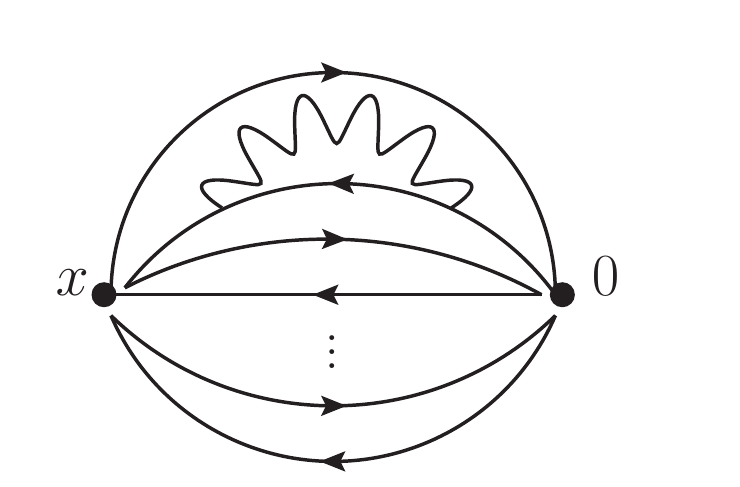}
 \caption{Example diagram for $\mathbb{L}$.\label{LEGEXAMPLE}}
 \end{center}
 \end{figure}

\subsubsection{Next-to-leading order}
\label{kfermNext}

\begin{figure}[t]
\begin{center}
\includegraphics[width = 1\textwidth]{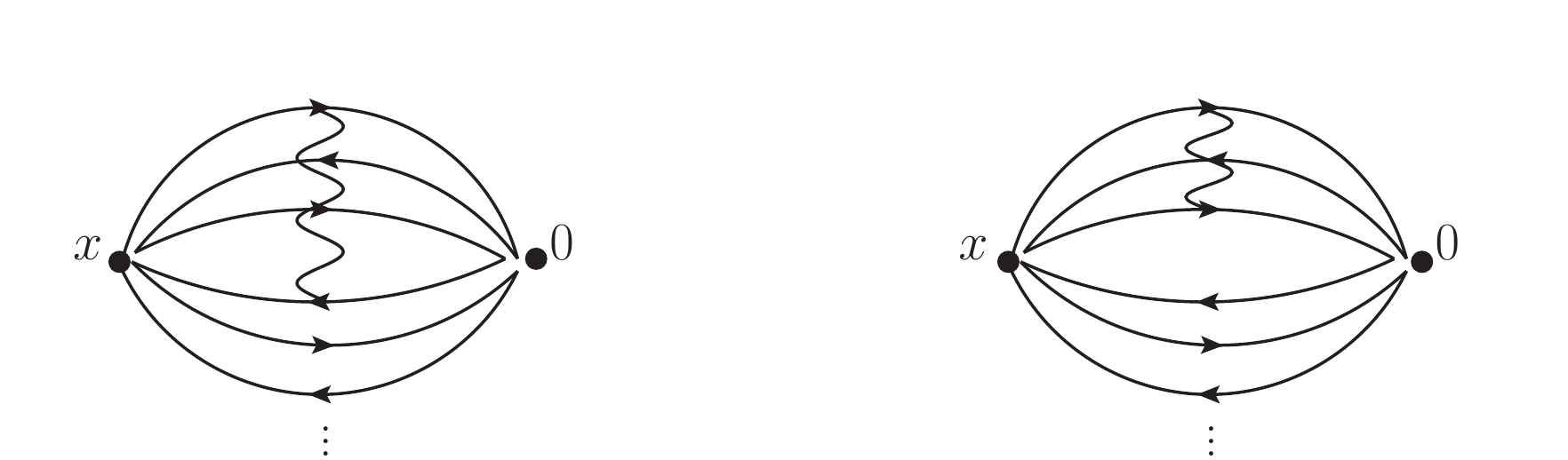}
 \caption{Diagrams for $\mathbb D_k(x)$ (left) and $\mathbb E_k(x)$ (right).\label{OPPOSITE}}
 \end{center}
 \end{figure}

For the next order in $1/N$, we should consider diagrams with one photon line.   In 
 \es{OOSub}{
  \langle\cO_n^I(x) \bar\cO_n^{C_{n,i}(I)}(0)\rangle_{(1)} \,,
 }
there are several possibilities for where to draw the photon line: 
 \begin{itemize}
  \item the photon line can connect a fermion line to itself.  Each such diagram gives
   \es{LegLowest}{
    \mathbb{L} = \langle\cO_n^I(x) \bar\cO_n^{C_{n,i}(I)}(0)\rangle_{(0)} \left[-\frac{\Delta_\psi^{(1)}}{N} \log (x^2 \Lambda^2) + O(\abs{x}^0) \right] \,.
   }
  There are $2j$ such diagrams for a permutation cycle of length $j$, for a total of $2n$ diagrams. See Figure~\ref{LEGEXAMPLE} for an example.
  \item the photon line can connect fermion lines belonging to different cycles of $C_{n,i}$.  These diagrams cancel in pairs---See Figure~\ref{CANCEL}.
  \item the photon line can connect distinct fermion lines of opposite types (one $G(x, 0)$ and one $G(0, x)$)  within the same cycle of $C_{n,i}$.  See the lefthand diagram in Figure~\ref{OPPOSITE}.  Let this cycle have length $k$. In position space, such a diagram is
  \es{DiagOpposite}{
 &{\mathbb{D}_k}(x) =-\left(\frac{(-1)^{k}2^{-1+\sum_ja_{ij}}}{(4\pi)^{2(n-k)}|x|^{4(n-k)}}\right)\text{sig}(C_{n,i}) \int d^3z\, d^3 w\,  D_{\mu\nu}(z, w)\\
&\times\tr \left[G(x, z) \gamma^\mu G(z, 0) (G(0, x) G(x, 0))^{k_1} G(0, w) \gamma^\nu G(w, x) (G(x, 0) G(0, x))^{k_2}  \right] \,,
}
  \begin{figure}[t]
\begin{center}
\includegraphics[width = 1\textwidth]{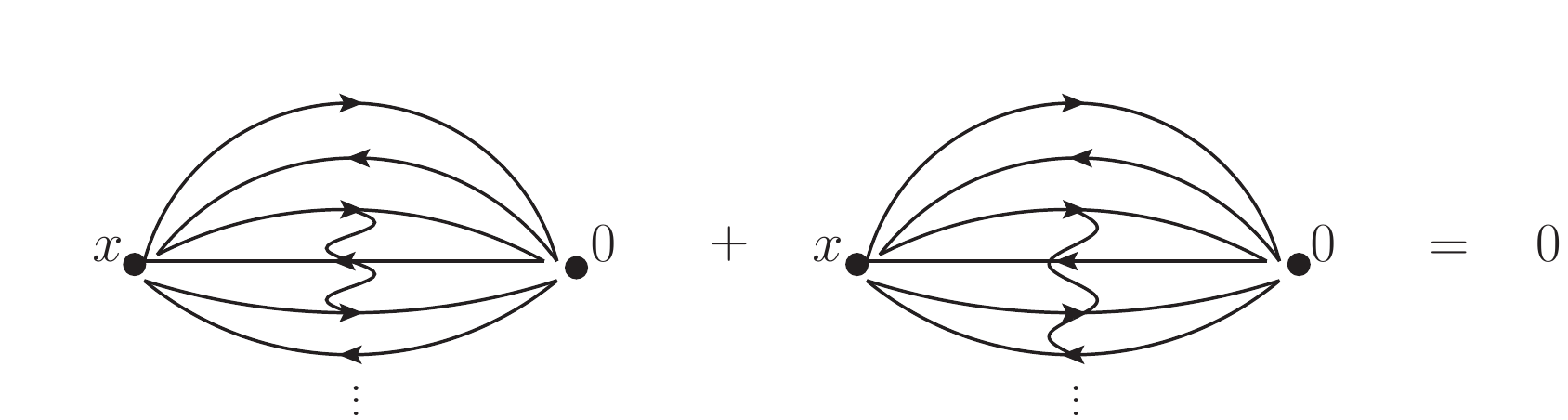}
 \caption{Diagrams that cancel in pairs.\label{CANCEL}}
 \end{center}
 \end{figure}
where the first term in parentheses comes from the cycles without photon lines, and the contribution we exhibited is that coming from the photon line.  The number of fermion propagators between those containing photon lines is $k_1$ and $k_2$, with $k_1 + k_2 = k-1$.
  \item the photon line can connect distinct fermion  lines of the same type (either both $G(x, 0)$ or both $G(0, x)$) within in the same cycle of $C_{n,i}$.  See the righthand diagram in Figure~\ref{OPPOSITE}. For instance, if the photon line connects two $G(x, 0)$'s in a cycle of length $k$ we have a contribution equal to
  \es{DiagSame}{
 &\mathbb{E}_k(x) = -\left(\frac{(-1)^{k}2^{-1+\sum_ja_{ij}}}{(4\pi)^{2(n-k)}|x|^{4(n-k)}}\right) \text{sig}(C_{n,i}) \int d^3z\, d^3 w\,  D_{\mu\nu}(z, w)\\
&\times\tr \left[G(x, z) \gamma^\mu G(z, 0) G(0, x) (G(x, 0) G(0, x))^{k_1} G(x,w) \gamma^\nu G(w, 0) G(0, x) (G(x, 0) G(0, x))^{k_2} \right] 
}
where again the first term in parentheses comes from the cycles without photon lines, and the contribution we exhibited is that coming from the photon line.  Here, $k_1 + k_2 = k-2$.
 \end{itemize} 

From $\mathbb{D}_k$ and $\mathbb{E}_k$ we have to extract the logarithmic divergence.  While these are very complicated diagrams and their full evaluation would be an onerous task, the extraction of the logarithmic divergence is quite easy, because it comes either from when $z$ and $w$ are both close to $x$ or to $0$.  Both limits give the same answer, so we can just take the limit where both $z$ and $w$ are close to $0$ and multiply the answer by $2$.  For $\mathbb{D}_k$, Eq.~\eqref{DiagOpposite} thus becomes 
  \es{DiagOppositeIntermediate}{
 &{\mathbb{D}_k}(x) \approx - \left(\frac{\text{sig}(C_{n,i}) 2^{\sum_ja_{ij}}}{(4\pi)^{2n}|x|^{4n}}\right)\int d^3z\, d^3 w\, D_{\mu\nu}(z, w)
 \tr \left[ \gamma^\mu G(z, 0)G(0, w) \gamma^\nu \right] \,,
}
where we used $G(x, 0) G(0, x) = - \frac{1}{(4 \pi)^2 x^4} I$.  The position space integral can be written in Fourier space as an integral over a Fourier momentum $q$:
\es{DiagOppositeIntermediate2}{
 \int d^3z\, d^3 w\, D_{\mu\nu}(z, w)
 \tr \left[ \gamma^\mu G(z, 0)G(0, w) \gamma^\nu \right]  &=\int \frac{d^3q}{(2\pi)^3}\tr\left[\gamma^\mu i\slashed q i\slashed q\gamma^\nu\right]\frac{D_{\mu\nu}(q)}{q^4} \,.
}
This expression can be seen to evaluate to $-8 (3 - \zeta) \log \Lambda^2 / (N \pi^2) $ after performing the required gamma matrix algebra and using the gauge field propagator in \eqref{Summary}.  Here, $\Lambda$ is the UV cutoff and it must appear inside the logarithm in the combination $\Lambda \abs{x}$.  Thus, $\mathbb{D}_k(x)$ evaluates to
\es{evenDiag2}{
 \mathbb{D}_k &=  \langle\cO_n^I(x) \bar\cO_n^{C_{n,i}(I)}(0)\rangle_{(0)} \left[\frac{8}{\pi^2N}(3-\zeta)\log\Lambda^2|x|^2
  + O(\abs{x}^0) \right]\,.
}

A similar strategy works for evaluating the logarithmic divergence in $\mathbb{E}_k$.  Taking the limits when $z, w$ go to $x$ or $0$, one obtains 
\es{oddDiagApprox}{
\mathbb{E}_k(x) 
  \approx \left(\frac{\text{sig}(C_{n,i}) 2^{\sum_ja_{ij}}}{(4\pi)^{2n}|x|^{4n}}\right)\int \frac{d^3q}{(2\pi)^3}\tr\left[\gamma^\mu i\slashed q\gamma^\nu i\slashed q\right]\frac{D_{\mu\nu}(q)}{q^4} \,.
}
The momentum space integral now gives $8 (1 + \zeta) \log \Lambda^2 / (N \pi^2) $, so in the end
\es{oddDiag}{
\mathbb{E}_k(x)
  = \langle\cO_n^I(x) \bar\cO_n^{C_{n,i}(I)}(0)\rangle_{(0)} \left[ \frac{8}{\pi^2N}(1+\zeta)\log\Lambda^2|x|^2 + O(\abs{x}^0) \right] \,.
}

Due to the various ways of placing the gauge propagator, there are $k^2$ diagrams that give $\mathbb{D}_k$ and $k(k-1)$ diagrams that give $\mathbb{E}_k$.  Along with the leg contributions, we have 
\es{DiagramSum}{
\langle\cO_n^I(x) \bar\cO_n^{C_{n,i}(I)}(0) = 
  \sum_{k=1}^n a_{ik} \left[ k^2 \mathbb{D}_k(x) + k(k-1) \mathbb{E}_k(x) + 2k \mathbb{L}(x) \right] \,.
}
Quite nicely, after plugging in \eqref{LegLowest}, \eqref{evenDiag2}, and \eqref{oddDiag} into \eqref{DiagramSum} one obtains an expression independent of $\zeta$:
\es{Mdiag}{
\langle\cO_n^I(x) \bar\cO_n^{C_{n,i}(I)}(0)\rangle & = \langle\cO_n^I(x) \bar\cO_n^{C_{n,i}(I)}(0)\rangle_{(0)}
 \left[ 
   \sum_{k=1}^n \frac{32k(3k-1) \log\Lambda^2|x|^2}{3 \pi^2N} + O(\abs{x}^0) \right] \,.
}
Using \eqref{TwoPointPerm} and \eqref{NOpleading}, we can write the ratio between the $1/N$ correction to the two-point function and the leading result as
\es{NOpsub}{
\frac{\langle\cO_n(x) \bar\cO_n(0)\rangle_{(1)}}{\langle\cO_n(x) \bar\cO_n(0)\rangle_{(0)}}
=   \frac{32 \sum_{C_{n,i}\in\text{Cl}(S_n)}|C_{n,i}|2^{\sum_ja_{ij}}\sum_{k}a_{ik} k(3k-1)}{3\pi^2N\sum_{C_{n,i}\in\text{Cl}(S_n)}|C_{n,i}|2^{\sum_ja_{ij}}}\log |x|^2\Lambda^2 + O(\abs{x}^0) \,.
}

The results of Section~\ref{GENERALANOMALOUS} then imply that the scaling dimension of ${\cal O}_n$ is 
\es{NOpFinal}{
\Delta^{}_n=2n- \frac{32}{3 \pi^2} \frac{ \sum_{C_{n,i}\in\text{Cl}(S_n)}|C_{n,i}|2^{\sum_ja_{ij}}\sum_{k}a_{ik} k(3k-1)}{\sum_{C_{n,i}\in\text{Cl}(S_n)}|C_{n,i}|2^{\sum_ja_{ij}}} \frac 1N +O(1/N^2)\,.
}
This expression can be evaluated for any $n$ using the data for the conjugacy classes of the permutation group.  When $n=1$, one has only one conjugacy class $C_1$ of size $\abs{C_1} = 1$ and $a_{11} = 1$;  it is easy to see that \eqref{NOpFinal} reduces to \eqref{Delta1Large}.

\subsection{Scaling dimension of ${\cal O}_n'$ and ${\cal O}_n''$}

\label{scalDim2}
We consider particular operators representing \eqref{kOp2} by choosing $i_k = k$:
\es{kOp2Def2}{
 \cO_n'&= \frac{1}{\sqrt{N}}\sum_{k=1}^N\psi^{[1}_{(\alpha_{1}}\dots\psi^{n}_{\alpha_{n}}\psi^{k]}_{\alpha_{n+1)}}\bar\psi_{[n+1}^{(\alpha_{1}}\dots\bar\psi_{2n}^{\alpha_{n}}\bar\psi_{k]}^{\alpha_{n+1})}\,,\\
 \cO_n'' &= \frac{\left(\bar\psi_i\psi^i\right)}{\sqrt{N}}\psi^{[1}_{(\alpha_1}\dots\psi^{n]}_{\alpha_n)}\bar\psi_{[n+1}^{(\alpha_1}\dots\bar\psi_{2n]}^{\alpha_n)} 
  = {\cal O}_n {\cal O}_0\,,
}
with ${\cal O}_n$ as in \eqref{NOpDef} and ${\cal O}_0$ as in Section~\ref{SINGLET}.  Since we have two operators that mix together, we must consider the matrix of 2-point functions
 \es{defM}{
  {\cal M}_n(x) = \begin{pmatrix}
   \langle {\cal O}_{n}'(x) \bar {\cal O}_{n}'(0) \rangle & \langle {\cal O}_{n}'(x) \bar {\cal O}_{n}''(0) \rangle \\
   \langle {\cal O}_{n}''(x) \bar {\cal O}_{n}'(0) \rangle & \langle {\cal O}_{n}''(x) \bar {\cal O}_{n}''(0) \rangle
  \end{pmatrix} \,,
 }
as in Section~\ref{GENERALANOMALOUS} and expand it in $1/N$.

Note that we can write
 \es{OpRewrite}{
  {\cal O}_n' &= 
    (-1)^{\frac{(n+1)(n+2)}{2}}\frac{1}{(n+1)!} \frac{1}{\sqrt{N}} \sum_{\sigma \in S_{n+1}}\sum_{k=1}^N \text{sig}(\sigma)  \tilde \cO_n^{(\sigma, k)}(x)\,, 
      \\
   \tilde \cO_n^{(\sigma, k)} &\equiv  \bar\psi_{n+1}\psi^{\sigma_k(1)}\dots\bar\psi_{2n}\psi^{\sigma_k(n)} 
    \bar \psi_k \psi^{\sigma_k(n+1)}\,,
 }
where $\sigma$ is a permutation of the set $\{1, \ldots, n+1\}$, and $\sigma_k = \pi_k \circ \sigma$, $\pi_k$ being the map $\pi_k(i) = i$ for $i=1, \ldots, n$ and $\pi_k(n+1) = k$.  This expression is somewhat similar to that for ${\cal O}_{n+1}$, which is an observation that will simplify some of our computations.

\subsubsection{$\langle {\cal O}_{n}''(x) \bar {\cal O}_{n}''(0) \rangle$}

The two-point function $\langle {\cal O}_{n}''(x) \bar {\cal O}_{n}''(0) \rangle$ is the simplest to calculate because it factorizes not just at leading order in $1/N$, but also at the first subleading order:
 \es{2pt22}{
 \langle {\cal O}_{n}''(x) \bar {\cal O}_{n}''(0) \rangle= \langle {\cal O}_{n}^{}(x) \bar {\cal O}_{n}^{}(0) \rangle \langle {\cal O}_0(x) \bar {\cal O}_0(0)\rangle+O(1/N^2) \,.
 }
The factorization at next-to-leading order is because the diagrams formed by photon lines between ${\cal O}_n$ and ${\cal O}_0$ all cancel in pairs. From \eqref{NOpsub} and \eqref{Delta0Large}, we have the following ratio of subleading to leading orders
 \es{2pt22Expanded}{
  \frac{\langle {\cal O}_{n}''(x) \bar {\cal O}_{n}''(0) \rangle_{(1)}}{\langle {\cal O}_{n}''(x) \bar {\cal O}_{n}''(0) \rangle_{(0)}} = \left[ \frac{ \sum_{C_{n,i}\in\text{Cl}(S_n)}|C_{n,i}|2^{\sum_ja_{ij}}\sum_{k}a_{ik} k(3k-1)}{\sum_{C_{n,i}\in\text{Cl}(S_n)}|C_{n,i}|2^{\sum_ja_{ij}}} - 4\right]\frac{32\log |x|^2\Lambda^2}{3 \pi^2 N} + O(\abs{x}^0) \,.
 }

\subsubsection{$\langle {\cal O}_{n}'(x) \bar {\cal O}_{n}'(0) \rangle$ and $\langle {\cal O}_{n}'(x) \bar {\cal O}_{n}''(0) \rangle$}

Calculating the two-point functions $\langle {\cal O}_{n}'(x) \bar {\cal O}_{n}'(0) \rangle$ and $\langle {\cal O}_{n}'(x) \bar {\cal O}_{n}''(0) \rangle$ is harder.  First, notice that
 \es{TwoPointPermNew}{
  \langle\cO_n''(x) \bar\cO_n'(0)\rangle
    &=  \frac{1}{(n+1)!} \frac{1}{N} \sum_{k=1}^N \sum_{\sigma \in S_{n+1}} 
   \text{sig}(\sigma) \left\langle \tilde \cO_n^{(I, k)}(x) \overline{\tilde \cO_n^{(\sigma, k)}}(0) \right\rangle \,, \\
  \langle\cO_n'(x) \bar\cO_n'(0)\rangle
    &=  \frac{1}{(n+1)!} \frac{1}{N} \sum_{k=1}^N \sum_{\sigma \in S_{n+1}} 
   \text{sig}(\sigma) \left\langle \left[ \frac{\tilde \cO_n^{(I, k)}(x)}{n+1} 
    + \frac{n\tilde \cO_n^{(\tilde I, k)}(x)}{n+1} \right] \overline{\tilde \cO_n^{(\sigma, k)}}(0) \right\rangle \,,
 } 
where $I$ is the identity permutation, and $\tilde I$ is the transposition that flips $n$ and $n+1$.  These expressions are valid to all orders in $1/N$.  The terms in the $k$ sum vanish if $k\leq 2n$, because then $\tilde {\cal O}_n^{(\sigma, k)}$ as defined in \eqref{OpRewrite} vanishes automatically.

At leading order in $1/N$, Eqs.~\eqref{TwoPointPermNew} simplify, and they become equal to the leading order two point function of the operator ${\cal O}_{n+1}$ that was studied in the previous section:  
\es{2pt20}{
\langle {\cal O}_{n}'(x) \bar {\cal O}_{n}'(0) \rangle_{(0)}=\langle {\cal O}_{n}'(x) \bar {\cal O}_{n}''(0) \rangle_{(0)}=\langle\cO^{}_{n+1}(x) \bar\cO^{}_{n+1}(0)\rangle_{(0)}\,.
}

  \begin{figure}[t]
\begin{center}
\includegraphics[width = 1\textwidth]{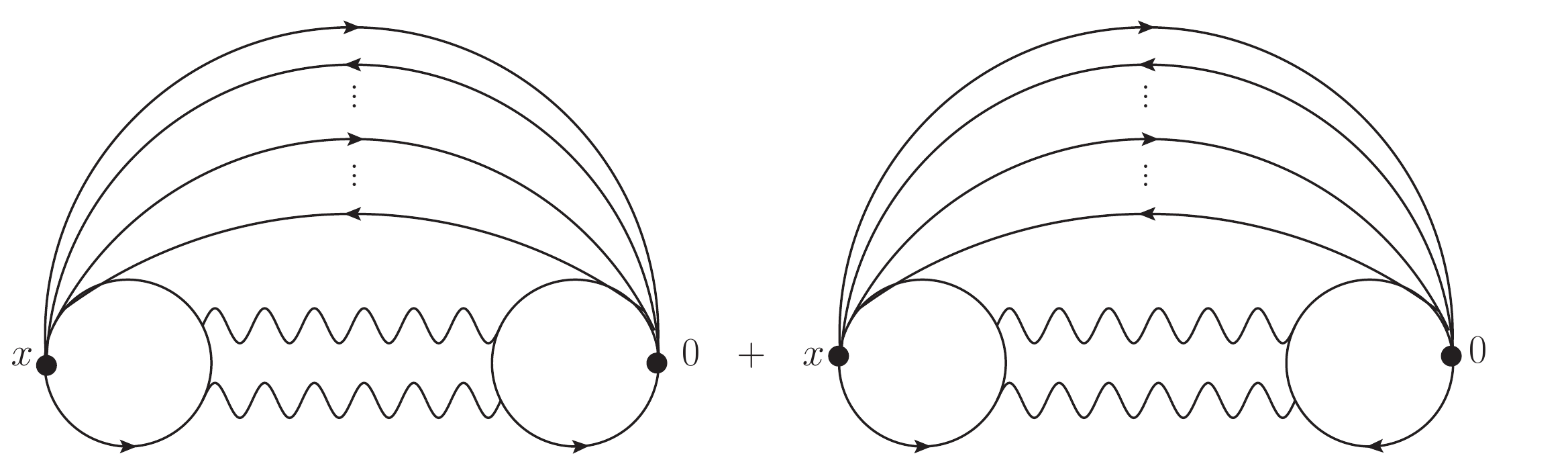}
 \caption{Additional diagrams for $\langle {\cal O}_{n}'(x) \bar {\cal O}_{n}''(0) \rangle_{(1)}$ and $\langle {\cal O}_{n}'(x) \bar {\cal O}_{n}'(0) \rangle_{(1)}$.\label{BUBBLES}}
 \end{center}
 \end{figure}

At sub-leading order, the two-point functions \eqref{2pt20} have the same diagrams as $\langle\cO^{}_{n+1}\bar\cO^{}_{n+1}\rangle_{(1)}$, but also differ from $\langle\cO^{}_{n+1}\bar\cO^{}_{n+1}\rangle_{(1)}$ due to the occurrence of additional diagrams where the $\psi_k$ and $\bar \psi^k$ belonging to either ${\cal O}_n'$ or ${\cal O}_n''$ are joined together by a fermion line.  See Figure~\ref{BUBBLES}.  These additional diagrams are similar to the last two diagrams in Figure~\ref{Delta0Diagrams}.  We thus obtain
\es{fermdiagram21}{
\langle {\cal O}_{n}'(x) \bar {\cal O}_{n}''(0) \rangle_{(1)}
&= \langle\cO^{}_{n+1}(x) \bar\cO^{}_{n+1}(0)\rangle_{(1)} \\
{}&+ \left(\frac{1-n/2}{n+1}\right)\langle {\cal O}_{n+1}(x) \bar {\cal O}_{n+1}(0) \rangle_{(0)}\left[ -\frac{64}{\pi^2N} \log|x|^2\Lambda^2 + O(\abs{x}^0) \right]\,,
} 
\es{fermdiagram22}{
\langle {\cal O}_{n}'(x) \bar {\cal O}_{n}'(0) \rangle_{(1)}
&= \langle\cO^{}_{n+1}(x) \bar\cO^{}_{n+1}(0)\rangle_{(1)} \\
{}&+ \left(\frac{1-n/2}{n+1}\right)^2\langle {\cal O}_{n+1}(x) \bar {\cal O}_{n+1}(0) \rangle_{(0)}\left[ -\frac{64}{\pi^2N} \log|x|^2\Lambda^2 + O(\abs{x}^0) \right]\,.
}

The $n$-dependent prefactors in the second lines of \eqref{fermdiagram21} and \eqref{fermdiagram22} can be understood as follows.   In the case where $\psi^k$ and $\bar \psi_k$ belong to ${\cal O}_n'$, we either have that $\psi^k$ can be contracted with $\bar\psi_k$, or $\psi^k$ and $\bar\psi_k$ can be part of a bigger cycle. Out of $(n+1)!$ total possibilities, the first case occurs $n!$ times, while the latter occurs $n!n$ times and has an extra factor of $-1/2$ relative to the first, because there is one fewer trace and permutation for this diagram. Summing both cases we find that whenever ${\cal O}_n'$ is involved we must include a factor of 
\es{prefactor}{
  \frac{n!-n!n/2}{(n+1)!} = \frac{1-n/2}{n+1} 
} 
relative to the $n=0$ case of the last two diagrams in Figure~\ref{Delta0Diagrams}.  For ${\cal O}_n''$ we do not need any extra factors.  Thus \eqref{fermdiagram21} contains one power of \eqref{prefactor} and \eqref{fermdiagram22} contains two powers of \eqref{prefactor}.

Gathering the previous results we can write down the ${\bf M}$ and ${\bf N}$ matrices defined in \eqref{DefineM}:
\es{k2Final}{
\bold{N}_{n}&=\begin{pmatrix}
A & A    \\
  A &B
  \end{pmatrix}\,,
\qquad
\bold{M}_{n}=\begin{pmatrix}
C& D    \\
  D &E
  \end{pmatrix}\,,
 }
where
 \es{ABDefs}{ 
A&=\frac{1}{(n+1)!}\sum_{C_{n+1,i}\in\text{Cl}(S_{n+1})}|C_{n+1,i}|\left(\frac{2^{\sum_{j=1}^{n+1} a_{ij}}}{(4\pi)^{2n+2}}\right)\,,\qquad B=\frac{2}{n!}\sum_{C_{n,i}\in\text{Cl}(S_n)}|C_{n,i}|\left(\frac{2^{\sum_{j=1}^n a_{ij}}}{(4\pi)^{2n+2}}\right)\,,\\
C&= \frac{4}{\pi^4(n+1)!N} \sum_{C_{n+1,i}\in\text{Cl}(S_{n+1})}|C_{n+1,i}|\left(\frac{2^{\sum_{j=1}^{n+1} a_{ij}}}{(4\pi)^{2n}}\right)\left[-\left(\frac{1-n/2}{n+1}\right)^2+\sum_{k=1}^{n+1}\frac{a_{ik} k(3k-1)}{6}\right]\,,\\
D&= \frac{4}{\pi^4(n+1)!N} \sum_{C_{n+1,i}\in\text{Cl}(S_{n+1})}|C_{n+1,i}|\left(\frac{2^{\sum_{j=1}^{n+1} a_{ij}}}{(4\pi)^{2n}}\right)\left[-\frac{1-n/2}{n+1}+\sum_{k=1}^{n+1}\frac{a_{ik} k(3k-1)}{6}\right]\,,\\
E&= \frac{8}{\pi^4n!N} \sum_{C_{n,i}\in\text{Cl}(S_{n})}|C_{n,i}|\left(\frac{2^{\sum_{j=1}^{n} a_{ij}}}{(4\pi)^{2n}}\right)\left[-\frac23+\sum_{k=1}^{n}\frac{a_{ik} k(3k-1)}{6}\right]\,.
}

 From this expression and \eqref{TotalScaling}, we can extract the anomalous dimensions by diagonalizing $ {\bf N}^{-1} {\bf M}$, which yields
\es{knOPs}{
\Delta'_{n,\pm}=&2n+2+\frac{2AD-CB-AE}{2A\left(A-B\right)}\pm\frac{\sqrt{\left(CB+AE-2AD\right)^2-4A\left(A-B\right)\left(D^2-CE\right)}}{2A\left(A-B\right)} \,.
}
Particular cases are given in \eqref{2462} in the Introduction.  

\section{The mixing of lowest parity-even $SU(N)$ singlets}
\label{4singlet}

We now consider the parity-even $SU(N)$ singlets.  At large $N$, all these operators are irrelevant.  As mentioned in the Introduction, it is important to estimate down to what value of $N$ this situation persists, because if a parity-even $SU(N)$ singlet becomes relevant, it can be generated during the RG flow and change the fate of the infrared physics.

At infinite $N$, there are two lowest-dimension parity even operators that mix in $1/N$ perturbation theory.  They are 
 \es{O12DefSinglets}{
  {\cal O}_1 = \frac{1}{N} (\bar \psi_i \psi^i) (\bar \psi_j \psi^j) \,, \qquad
   {\cal O}_2 = \frac{N}{4} F_{\mu\nu} F^{\mu\nu} \,,
 }
where the factors of $N$ have been chosen such that the two-point functions of these operators scale as $N^0$ at large $N$.  Both operators in \eqref{O12DefSinglets} are real, and their scaling dimension is $\Delta_0 = 4$ at $N=\infty$.

\begin{figure}[t]
\begin{center}
\includegraphics[width = 1\textwidth]{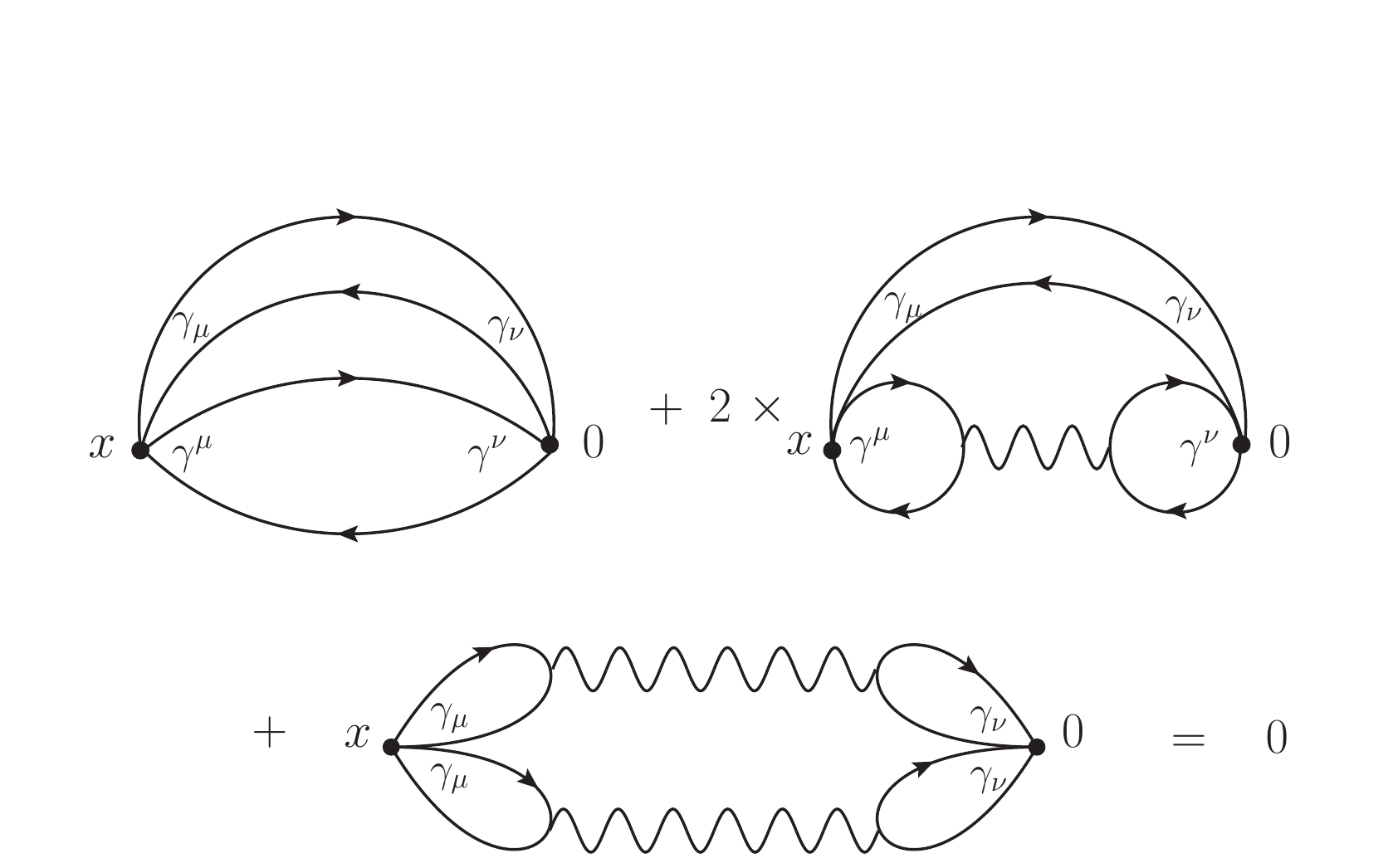}
 \caption{Cancelation of diagrams contributing to $\langle {\cal O}_3(x) {\cal O}_3 (0) \rangle$ at order $N^0$. \label{fig:diagCancel}}
 \end{center}
 \end{figure}

Before we start calculating the mixing between these two operators, let us explain why there are only two such operators.   The counting argument of Section~\ref{numOps} implies that there are actually two linearly-independent four-fermion operators that are $SU(N)$ singlets.  They can be taken to be ${\cal O}_1$ and ${\cal O}_3 = \frac{1}{N} (\bar \psi_i  \gamma_\mu \psi^i) (\bar \psi_j \gamma^\mu \psi^j)$.  However, ${\cal O}_3$ is proportional to the large $N$ equation of motion operator $E_\mu = \bar \psi_i \gamma_\mu \psi^i = 0$ of the large $N$ theory obtained by varying the action with respect to $A_\mu$.  As such, ${\cal O}_3$ does not contribute to the matrix of two-point functions.  Indeed, it can be checked that at separated points we have $\langle {\cal O}_3(x) {\cal O}_3 (0) \rangle = 0$.  For instance, at order $N^0$, the diagrams in Figure~\ref{fig:diagCancel} can be seen to cancel exactly.\footnote{We thank Mark Mezei for discussions on this issue.}

Let us thus focus on the matrix of two-point functions of ${\cal O}_i$, with $i=1, 2$ and write it at large $N$ in the form given in Section~\ref{GENERALANOMALOUS}.  In particular, let us compute the matrices ${\bf M}$ and ${\bf N}$ defined in \eqref{DefineM}.

At leading order in $N$, we have
 \es{psi4Leading}{
  \langle {\cal O}_1(x) {{\cal O}}_1(0) \rangle_{(0)} = 2 \tr \left[ G(x, 0) G(0, x) \right] \tr \left[ G(x, 0) G(0, x) \right] = \frac{1}{32 \pi^4 \abs{x}^8} \,.
 }
Rewriting ${\cal O}_2$ as
  \es{O2Rewrite}{
   {\cal O}_2 = \frac{N}{2} \left[ \partial_\mu A_\nu \partial^\mu A^\nu - \partial_\mu A_\nu \partial^\nu A^\mu \right] \,,
  }
and using the gauge field propagator $D_{\mu\nu}$ in \eqref{Summary} gives 
 \es{O2TwoPoint}{
  \langle {\cal O}_2 (x) {{\cal O}}_2 (0) \rangle_{(0)} 
     &= 3 \frac{512 }{\pi^4 \abs{x}^8}  \,.
 }
Since at order $N^0$, the two-point function $\langle {\cal O}_1(x) {{\cal O}}_2(0) \rangle_{(0)}$ vanishes, the matrix ${\bf N}$ defined in \eqref{DefineM} is
 \es{M0Singlet}{
  {\bf N} = \frac{1}{\pi^4 } 
  \begin{pmatrix}
   \frac{1}{32}  & 0 \\
   0 & 3 \times 512
  \end{pmatrix} \,.
 }

In order to compute ${\bf M}$, it is natural to think of each ${\cal O}_1$ as a composite between $\bar \psi_i \psi^i$ and $\bar \psi_j \psi^j$, and of ${\cal O}_2$ as the composite between $F_{\mu\nu}$ and $F^{\mu\nu}$.  There are many diagrams that contribute to ${\bf M}$ but they can be split into diagrams referred to as leg corrections coming from each of the factors of the composites as well as diagrams referred to as vertex corrections that mix together the two factors.

The leg correction diagrams have already been computed.  Since the operator $\bar \psi_i \psi^i$ acquires an anomalous dimension given by $128/(3 \pi^2 N) + O(1/N^2)$ (see \eqref{DeltaSinglet}) and $F_{\mu\nu}$ has no anomalous dimension, we have that the leg contribution to ${\bf M}$ is
 \es{M1Leg}{
  {\bf M}^\text{leg} = \frac{1}{\pi^4} 
  \begin{pmatrix}
   \frac{1}{32} \times \frac{256}{3 \pi^2 } & 0 \\
   0 & 0
  \end{pmatrix} \,.
 }

\begin{figure}[t]
\begin{center}
\includegraphics[width = 1\textwidth]{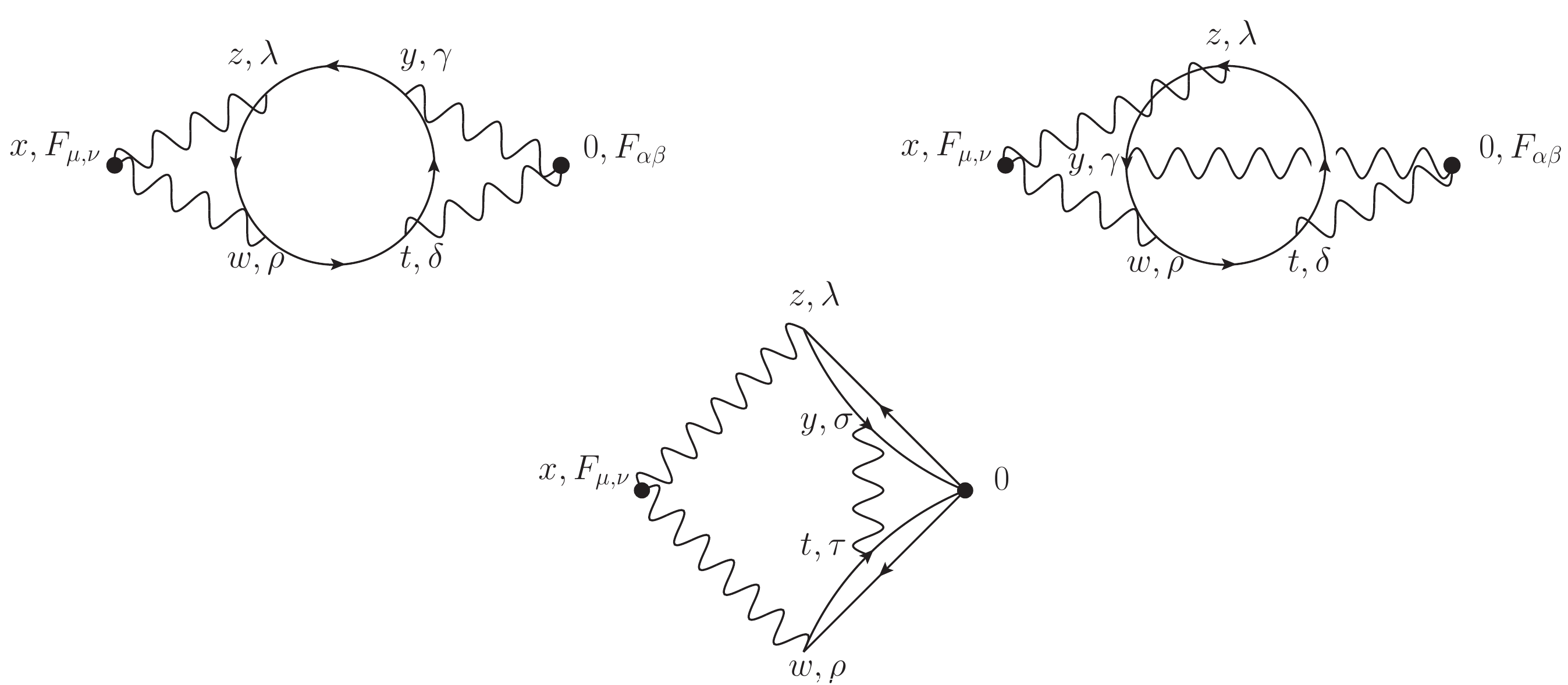}
 \caption{Diagrams that contribute to the mixing of the operators ${\cal O}_1$ and ${\cal O}_2$ defined in \eqref{O12DefSinglets}. \label{fig:diags}}
 \end{center}
 \end{figure}

Let us now compute the leading vertex corrections.  The two-point function $\langle {\cal O}_1(x) {\bar{\cal O}}_1(0) \rangle$ does not receive any such corrections.  (The diagrams cancel in pairs as in Figure~\ref{CANCEL}.)  On the other hand, $\langle {\cal O}_1(x) {\bar{\cal O}}_2(0) \rangle = \langle {\cal O}_2(x) {\bar{\cal O}}_1(0) \rangle$ and $\langle {\cal O}_2(x) {\bar{\cal O}}_2(0) \rangle$ might.  

Let's start with $\langle {\cal O}_1(x) {\bar{\cal O}}_2(0) \rangle$ (bottom diagram in Figure~\ref{fig:diags}).  In position space, this diagram can be written as
 \es{O21VertexSimp}{
  \langle {\cal O}_2(x) {\bar{\cal O}}_1(0) \rangle_\text{vertex} &=  4 N^2 \int d^3z \, d^3 w\, d^3 y \, d^3 t \,
    \partial_\mu D_{\nu \lambda}(x, z)
   \left(- \partial^\mu D_{\rho}{}^\nu(w, x)
    + \partial^\nu D_{\rho}{}^{\mu}(w, x) \right)    \\
   &\times    D_{\sigma\tau}(y, t) \tr \left[G(0, z) \gamma^\lambda G(z, y) \gamma^\sigma G(y, 0) \right]  
      \tr \left[G(0, t) \gamma^\tau G(t, w) \gamma^\rho G(w, 0) \right] \,.
 }
As in the previous sections, we are content with extracting only the logarithmic divergence of this diagram, leaving its full evaluation to future work.  The log divergences come from regions where $z$, $w$, $y$, $t$ are either close to $x$ or to $0$.  The first one is
 \es{O21Vertex1}{
  \langle {\cal O}_2(x) {\bar{\cal O}}_1(0) \rangle_\text{vertex}^{(x)} &\approx  4N^2  \int d^3z \, d^3 w\, d^3 y \, d^3 t \,
   \partial_\mu D_{\nu \lambda}(x, z)
     \left(- \partial^\mu D_{\rho}{}^\nu(w, x)
    + \partial^\nu D_{\rho}{}^{\mu}(w, x) \right)   \\
   &\times    D_{\sigma\tau}(y, t)  \tr \left[G(0, x) \gamma^\lambda G(z, y) \gamma^\sigma G(x, 0) \right] 
     \tr \left[G(0, x) \gamma^\tau G(t, w) \gamma^\rho G(x, 0) \right] \,.
 }
This is just a one-loop diagram, written in Fourier space as
 \es{O21Vertex1Again}{
  \langle {\cal O}_2(x) {\bar{\cal O}}_1(0) \rangle_\text{vertex}^{(x)} &\approx  -4N^2
   \int \frac{d^3 q}{(2 \pi)^3}  q_\mu D_{\nu \lambda}(q)  \left( q^\mu D_{\rho}{}^\nu(q)
    - q^\nu D_{\rho}{}^{\mu}(q) \right)   \\
   &\times    D_{\sigma\tau}(q) \tr \left[(-\slashed{x}) \gamma^\lambda i \slashed{q} \gamma^\sigma \slashed{x} \right] 
      \tr \left[(-\slashed{x}) \gamma^\tau i \slashed{q} \gamma^\rho \slashed{x} \right]
       \frac{1}{(4 \pi)^4 \abs{x}^{12} \abs{q}^4 } \,.
 }
This expression evaluates to 
 \es{O21Vert1Final}{
   \langle {\cal O}_2(x) {\bar{\cal O}}_1(0) \rangle_\text{vertex}^{(x)} &\approx  \frac{512}{N \pi^4 \abs{x}^8} \int \frac{d^3 q}{(2 \pi)^3} \frac{1}{\abs{q}^3} 
     =  \frac{128}{N \pi^6 \abs{x}^8}  \log (\Lambda^2 \abs{x}^2) \,.
 }

To evaluate the contribution from when $z$, $w$, $y$, $t$ are close to $0$ in \eqref{O21VertexSimp} one has to be more careful.  To obtain the log divergence, one has to expand the $D$'s in the first line of \eqref{O21VertexSimp} to linear order in $z$ and $w$ as these quantities tend to zero:
 \es{O21VertexSimp0}{
  \langle {\cal O}_2(x) {\bar{\cal O}}_1(0) \rangle_\text{vertex}^{(0)} &\approx  4 N^2
    \partial_\mu \partial_\alpha D_{\nu \lambda}(x)
   \partial_\beta \left( \partial^\mu D_{\rho}{}^\nu(x)
    - \partial^\nu D_{\rho}{}^{\mu}(x) \right)   F^{\alpha \lambda \beta \rho} \,, 
  }
where
 \es{FDef}{    
   F^{\alpha \lambda \beta \rho}(x)&=   \int d^3z \, d^3 w\, d^3 y \, d^3 t \,    D_{\sigma\tau}(y, t)\,  z^\alpha w^\beta   \\
   &\times
      \tr \left[G(0, z) \gamma^\lambda G(z, y) \gamma^\sigma G(y, 0) \right] \tr \left[G(0, t) \gamma^\tau G(t, w) \gamma^\rho G(w, 0) \right] \,.
 }
(The terms proportional to $z^\alpha z^\beta$ and $w^\alpha w^\beta$ in the expansion give vanishing contribution to the final answer and can be dropped.)  In \eqref{FDef}, the $x$ dependence appears only implicitly as $\Lambda \abs{x}$, $\Lambda$ being the UV cutoff.   This is a 3-loop diagram
\es{FMom}{    
   F^{\alpha \lambda \beta \rho}(x)&=   \int \frac{d^3q}{(2 \pi)^3} \frac{d^3r}{(2 \pi)^3} \frac{d^3s}{(2 \pi)^3} \frac{16}{N \abs{q}} \left( \delta_{\sigma \tau} -\zeta \frac{q_\sigma q_\tau}{q^2} \right)  
     \frac{1}{\abs{r}^2 \abs{s}^2 \abs{r + q}^2 \abs{s + q}^2}  \\
   &\times
      \tr \left[\left( -\frac{\partial}{\partial r_\alpha} i \frac{\slashed{r}}{\abs{r}^2} \right) \gamma^\lambda i \slashed{r} \gamma^\sigma i (\slashed{r} + \slashed{q}) \right] \tr \left[i (\slashed{s} + \slashed{q}) \gamma^\tau i \slashed{s} \gamma^\rho \left( \frac{\partial}{\partial s_\beta} i \frac{\slashed{s}}{\abs{s}^2} \right) \right]     
       \,.
 } 
It can be evaluated by first performing the $r$ and $s$ integrals, which give
 \es{FMomFinal}{
  F^{\alpha \lambda\beta \rho}(x)&\approx  
     \frac{\delta^{\alpha \beta} \delta^{\lambda \rho} - \delta^{\alpha \rho} \delta^{\beta \lambda} }{96 \pi^2 N } \log (\Lambda^2 \abs{x}^2)  \,.
 }
Going back to \eqref{O21VertexSimp0}, we have
 \es{O21VertexSimp0Again}{
  \langle {\cal O}_2(x) {\cal O}_1(0) \rangle_\text{vertex}^{(0)} &\approx  4 N^2
    \partial_\mu \partial_\alpha D_{\nu \lambda}(x)
   \partial_\beta \left( \partial^\mu D_{\rho}{}^\nu(x)
    - \partial^\nu D_{\rho}{}^{\mu}(x) \right)   \frac{\delta^{\alpha \beta} \delta^{\lambda \rho} - \delta^{\alpha \rho} \delta^{\beta \lambda}  }{96 \pi^2 N } \log (\Lambda^2 \abs{x}^2)  \,.
  }
Comparing with \eqref{O2TwoPoint}, we obtain
 \es{O21VertexSimpAgain}{
  \langle {\cal O}_2(x) {\cal O}_1(0) \rangle_\text{vertex}^{(0)} &\approx  
    \langle {\cal O}_2(x) {\cal O}_2(0) \rangle_{(0)}
      \frac{1 }{12 \pi^2 N } \log (\Lambda^2 \abs{x}^2)
       = \frac{1}{\pi^4 \abs{x}^8} 
      \frac{128 }{ \pi^2 N } \log (\Lambda^2 \abs{x}^2) \,.
 }

Adding up \eqref{O21Vert1Final} and \eqref{O21VertexSimpAgain}, we see that 
 \es{O2O1VertexTotal}{
   \langle {\cal O}_2(x) {\cal O}_1(0) \rangle_\text{vertex} \approx \frac{1}{\pi^4 \abs{x}^8} 
     \frac{256}{\pi^2 N} \log \left( \Lambda^2 \abs{x}^2 \right)  \,.
 }

Next, let's move on to $\langle {\cal O}_2(x) {\cal O}_2(0) \rangle$.  The diagrams that contribute are the top two diagrams in Figure~\ref{fig:diags}.  They are
 \es{O2O2VertexSimp}{
  \langle {\cal O}_2(x) {\cal O}_2(0) \rangle_\text{vertex}^{(1)}
   &= -N^3 \int d^3z\, d^3w\, d^3y\, d^3 t\, \partial_\mu D_{\nu \lambda}(x, z)
   \left(- \partial^\mu D_{\rho}{}^\nu(w, x)
    + \partial^\nu D_{\rho}{}^{\mu}(w, x) \right) \\
  &\times   \partial_\alpha D_{\beta \gamma}(0, y)
   \left(- \partial^\alpha D_{\delta}{}^\beta(t, 0)
    + \partial^\beta D_{\delta}{}^{\alpha}(t, 0) \right) \\
   &\times \Biggl[ \tr \left[ G(z, y) \gamma^\gamma G(y, t) \gamma^\delta G(t, w) \gamma^\rho G(w, z) \gamma^\lambda \right]  \\
   &+ \frac 12 \tr \left[ G(z, y) \gamma^\gamma G(y, w) \gamma^\rho G(w, t) \gamma^\delta G(t, z) \gamma^\lambda \right] \Biggr] \,,
 }
where the third and fourth lines represent the contributions of the top left and top right diagrams of Figure~\ref{fig:diags}, respectively.  There are potential divergences from when the intermediate points are close to $x$ and from when they are close to $0$, giving equal contributions, so we can consider the case when they're close to $0$ and multiply the answer by $2$.  Expanding to quadratic order in $z$ and $w$, we obtain
 \es{O2O2VertexExpansion}{
  \langle {\cal O}_2(x) {\cal O}_2(0) \rangle_\text{vertex}
   &\approx -2 N^3 \biggl[ \left[ \partial_\sigma \partial_\mu D_{\nu \lambda}(x) \right]
   \partial_\tau \left( \partial^\mu D_{\rho}{}^\nu(x)
    - \partial^\nu D_{\rho}{}^{\mu}(x) \right) H_1^{\sigma \lambda \tau \rho}(x) \\
  &+\left[ \partial_\sigma \partial_\tau  \partial_\mu D_{\nu \lambda}(x) \right]
  \left( \partial^\mu D_{\rho}{}^\nu(x)
    - \partial^\nu D_{\rho}{}^{\mu}(x) \right) H_2^{\sigma \lambda \tau \rho}(x) \\
  &+ \left[  \partial_\mu D_{\nu \lambda}(x) \right]
  \partial_\sigma \partial_\tau  \left( \partial^\mu D_{\rho}{}^\nu(x)
    - \partial^\nu D_{\rho}{}^{\mu}(x) \right) H_3^{\sigma \lambda \tau \rho}(x)  \biggr] \,,
 }
where 
 \es{NewFDef}{    
  H_i^{\sigma \lambda \tau \rho}(x) &=  \int d^3z\, d^3w\, d^3y\, d^3 t\, 
    \partial_\alpha D_{\beta \gamma}(0, y)
   \left(- \partial^\alpha D_{\delta}{}^\beta(t, 0)
    + \partial^\beta D_{\delta}{}^{\alpha}(t, 0) \right) \\
  &\times \Biggl[ \tr \left[ G(z, y) \gamma^\gamma G(y, t) \gamma^\delta G(t, w) \gamma^\rho G(w, z) \gamma^\lambda \right]  \\
   &+ \frac 12 \tr \left[ G(z, y) \gamma^\gamma G(y, w) \gamma^\rho G(w, t) \gamma^\delta G(t, z) \gamma^\lambda \right] \Biggr] \\  
  &\times \begin{cases}
    z^\sigma w^\tau \qquad  \text{if $i=1$} \,, \\
    -z^\sigma z^\tau  \qquad  \text{if $i=2$} \,, \\
     -w^\sigma w^\tau  \qquad  \text{if $i=3$} \,, 
   \end{cases}
 }
and the factor of $2$ in \eqref{O2O2VertexExpansion} is precisely because we're also accounting from the divergent contribution from the regime where the internal points are close to $x$.   The $x$ dependence of $H^{\sigma \lambda \tau \rho}_i$ is again implicit and appears only through $\Lambda \abs{x}$, where $\Lambda$ is the UV cutoff.   In momentum space, $H_1$ can be written as
 \es{FNewMom}{
 &H_1^{\sigma \lambda \tau \rho} = -\int \frac{d^3q}{(2 \pi)^3} \frac{d^3r}{(2 \pi)^3}
  \left[ \frac{\partial}{\partial q_\sigma} \left( q_\alpha D_{\beta \gamma}(q) \right) \right]
  \left[ \frac{\partial}{\partial q_\tau} \left(q^\alpha D_{\delta}{}^\beta(q) - q^\beta D_{\delta}{}^\alpha(q) \right) \right] \\
   &\times \Biggl[ \tr \left[G(r) \gamma^\gamma G(r + q) \gamma^\delta G(r) \gamma^\rho G(r) \gamma^\lambda \right] + \frac 12 \tr \left[G(r) \gamma^\gamma G(r + q) \gamma^\rho G(r + q) \gamma^\delta G(r) \gamma^\lambda \right]  \Biggr] \,.
 }
The expressions for $H_2$ and $H_3$ differ from \eqref{FNewMom} only in the placement of the derivatives $\partial_\sigma$ and $\partial_\tau$.  It is tedious but straightforward to perform the $r$ integral first and then the $q$ integral using the formulas in \cite{Smirnov:2012gma}.  The result is that these integrals do not have logarithmic divergences.

Putting everything together, we have that
 \es{GotM1}{
  {\bf M}^\text{vertex} = \begin{pmatrix}
   0 & \frac{256}{\pi^6} \\
   \frac{256}{\pi^6} & 0 
  \end{pmatrix} \,.
 }
Combining with \eqref{M0Singlet} with \eqref{GotM1} and \eqref{M1Leg}, we can write down the anomalous dimension matrix 
 \es{AnomSinglet}{
  {\bf N}^{-1} {\bf M} = {\bf N}^{-1} \left({\bf M}^\text{leg}+{\bf M}^\text{vertex}  \right)
   = \begin{pmatrix}
    \frac{256}{3 \pi^2} & \frac{8192}{\pi^2 } \\
    \frac{1}{6\pi^2 } & 0  
   \end{pmatrix}  \,.
 }
From the eigenvalues of this matrix, 
 \es{Evalues}{
  \frac{64 (2 \pm \sqrt{7}  ) }{3 \pi^2} \,,
 }
we conclude that the two parity even $SU(N)$ singlet operators have scaling dimensions given in \eqref{ScalingSinglet}.

\section*{Acknowledgments}

We thank Simone Giombi, John Gracey, Igor Klebanov, Mark Mezei, Grisha Tarnopolsky, and Cenke Xu for useful discussions.  This work was supported in part by the US NSF under Grant No.~PHY-1418069.

\appendix
\section{Useful Integrals and Fourier transforms in 3d}
\label{Appendix}

The following Fourier transform formulas are useful for the computations in this paper:
 \es{FT4}{
  \int d^3 x\, \frac{e^{i p \cdot x}}{\abs{x}^2} &= \frac{2 \pi^2}{\abs{p}} \,, \\
  \int d^3 x \, \frac{e^{i p \cdot x}}{\abs{x}^4} &= -\pi^2 \abs{p} \,, \\
  \int d^3 x \, \frac{e^{i p \cdot x}}{\abs{x}^6} &= \frac{\pi^2}{12} \abs{p}^3 \,, \\
  \int d^3 x \, \frac{e^{i p \cdot x}}{\abs{x}^8} &= -\frac{\pi^2}{360} \abs{p}^5 \,.
 }

Since 
 \es{Derp}{
   \partial_\mu \partial_\nu \abs{p} = \frac{1}{\abs{p}} \left( \delta_{\mu\nu} - \frac{p_\mu p_\nu}{p^2} \right) 
 } 
we have
 \es{IFTD}{
  \int d^3 x\, e^{i p \cdot x} \left[A \frac{\delta_{\mu\nu}}{\abs{x}^2} +  B \frac{x_\mu x_\nu}{\abs{x}^4} \right] = \pi^2 \frac{1}{\abs{p}} \left( (2A+B) \delta_{\mu\nu} - B \frac{p_\mu p_\nu}{p^2} \right)  \,.
 } 
To find the position space representation of the gauge propagator, we need
 \es{ABGauge}{
  A = \frac{8(1 - \zeta)}{\pi^2 N} \,, \qquad B = \frac{16 \zeta}{\pi^2 N} \,.
 }

\bibliographystyle{ssg}
\bibliography{largeN}

\end{document}